\documentclass[aps,twocolumn,superscriptaddress,flushbottom]{revtex4-1}
\usepackage[USenglish]{babel} 
\usepackage[T1]{fontenc}

\usepackage{amsmath}    
\usepackage{amssymb}	
\usepackage{relsize}
\usepackage{graphicx}   
\usepackage{verbatim}   
\usepackage{color}      
\usepackage{subfigure}  
\usepackage{hyperref}   
\usepackage{soul}
\usepackage{natbib}
\usepackage{setspace}
\usepackage{placeins}

\definecolor{light-blue}{rgb}{0.859375,0.898438,0.945317}
\definecolor{light-red}{rgb}{0.945,0.859,0.8554}
\definecolor{bright-blue}{rgb}{0,0.6875,1}
\definecolor{purple}{rgb}{.86,.4,.95}

\begin{document}

\title{Fast, high-resolution surface potential measurements in air\\ with heterodyne Kelvin probe force microscopy}
\author{Joseph L. Garrett}
	\address{ University of Maryland Department of Physics, College Park, MD 20742, USA}
	\address{ Institute for Research in Electronics and Applied Physics, College Park, MD 20742}
\author{Jeremy N. Munday}
	
	\address{ University of Maryland Department of Physics, College Park, MD 20742, USA}
	\address{ Institute for Research in Electronics and Applied Physics, College Park, MD 20742}
	\address{ Department of Electrical and Computer Engineering, College Park, MD 20742}

\begin{abstract}
Kelvin probe force microscopy (KPFM) adapts an atomic force microscope to measure electric potential on surfaces at nanometer length scales. Here we demonstrate that Heterodyne-KPFM enables scan rates of several frames per minute in air, and concurrently maintains spatial resolution and voltage sensitivity comparable to frequency-modulation KPFM, the current spatial resolution standard. Two common classes of topography-coupled artifacts are shown to be avoidable with H-KPFM. A second implementation of H-KPFM is also introduced, in which the voltage signal is amplified by the first cantilever resonance for enhanced sensitivity. The enhanced temporal resolution of H-KPFM can enable the imaging of many dynamic processes, such as such as electrochromic switching, phase transitions, and device degredation (battery, solar, {\it etc}.), which take place over seconds to minutes and involve changes in electric potential at nanometer lengths.
\end{abstract}

\maketitle 
\section{Introduction}
		The original amplitude-modulation Kelvin probe force microscopy (AM-KPFM) method\cite{Nonnenmacher1991} has been used in numerous studies investigating nanoscale phenomenon including: potential contrast between metals\cite{{Nonnenmacher1992}}, components of integrated circuits\cite{Weaver1991}, semiconductor doping\cite{Henning1995a}, pn junctions\cite{Kikukawa1995}, self-assembled monolayers\cite{Lu1999}, Langmuir films\cite{Reitzel2001}, crystal orientation of metals\cite{Gaillard2006}, and biomolecular binding to DNA\cite{Sinensky2007}. Developments such as lift mode\cite{Jacobs1997} alleviated problems with adhesion and allowed the investigation of softer surfaces\cite{Lee2000,Palermo2006}. AM-KPFM may seem suited for fast measurements, as it can operate quickly, and scan speeds of over 1 mm/s have been reported\cite{Sinensky2007}; however, in AM-KPFM, the voltage contrast is typically only a qualitative representation of the surface potential due to an averaging effect of the cantilever, the stray capacitance effect\cite{Jacobs1997,Colchero2001,Koley2001,Ma2013}. Moreover, AM-KPFM is susceptible to a class of artifacts that originate from interfering signals and appear in traditional KPFM measurements as topographical coupling\cite{Diesinger2008,Melin2011,Polak2014,Barbet2014}.
		
		The development of Frequency-Modulation (FM) KPFM\cite{Kitamura1998} improved spatial resolution and repeatability\cite{Glatzel2003, Zerweck2005, Panchal2013} and has been used to quantitatively compare nanoscale potentials with macroscopic work functions on both semiconductors\cite{Sommerhalter1999} and graphene\cite{Panchal2013}, to identify semiconductor crystal orientations\cite{Sadewasser2002}, to characterize lipid self-organization\cite{Leonenko2007}, to quantify band bending at grain boundaries\cite{Sadewasser2003}, to study charge transport and trapping in quantum dots\cite{Zhang2015}, and to investigate the charge distribution at sub-molecular and atomic length scales\cite{Mohn2012,Gross2009}. However, dynamics are difficult to measure with FM-KPFM because of its slow scan speeds---the result of potential and topographic feedback loops detected near the same cantilever resonance, limiting detection bandwidth\cite{Glatzel2003,Zerweck2005}. 
		
		Techniques that try to couple the repeatability and spatial resolution of FM-KPFM with enhanced time resolution include time resolved electrostatic force microscopy and pump-probe KPFM, which both probe the dynamic response to an impulse point-by-point\cite{Coffey2006a,Murawski2015}, and open loop (OL) KPFM techniques, which eliminate the KPFM voltage feedback loop\cite{Takeuchi2007,Collins2013, Borgani2014a}. However, not every dynamic process is caused by an impulse, and the typical scan speed with high-resolution open-loop techniques is about 1 $\mu$m/s\cite{Takeuchi2007,Borgani2014a}, slower than AM-KPFM.
		
		Operation in air is necessary to study biological molecules such as lipids and DNA\cite{Moores2010,Sinensky2007} and to study solar cell properties such as open-circuit voltage and degradation in realistic operation conditions\cite{Tennyson,Sengupta2011}. However, developments of KPFM have often focused on operation in vacuum\cite{Kitamura1998,Sugawara2012}. In air, challenges such as vastly lower Q factors, which reduce sensitivity, and a thin adhesive water layer must be overcome\cite{Wastl2013a}.  
		
		A recent technique, Heterodyne (H) KPFM, operates similarly to FM-KFPM but separates the topography and voltage signals by hundreds of kHz\cite{Sugawara2012}. Originally, in vacuum, the separation was utilized to increase the voltage sensitivity through amplification by the second cantilever eigenmode, while maintaining spatial resolution equal to FM-KPFM\cite{Sugawara2012,Ma2013a}. Measurements in vacuum show that H-KPFM, like FM-KPFM, avoids the stray capacitance artifact that affects AM-KPFM\cite{Ma2013}.
		
		Here we demonstrate that H-KPFM combines the repeatability and spatial resolution of FM-KPFM with scan speeds of up to 32 $\mu$m/s (1x1 $\mu$m, 256$\times$256 pixels, 16 s, trace and retrace). Moreover, H-KPFM achieves its time resolution without requiring an impulse. We show that it is not susceptible to several topographical artifacts that hinder the other KPFM methods. We demonstrate that it is compatible with lift mode. A second implementation of H-KFPM is also introduced, in which the topography is detected with the second cantilever resonance and the voltage with the first, for additional voltage sensitivity. The temporal resolution, voltage contrast, and spatial resolution of H-KPFM are each compared to those of both FM- and AM-KPFM. It is deduced that H-KPFM improves upon the spatial resolution of AM-KPFM and improves upon the scan speed of FM-KPFM, resulting in a new technique with improved performance in ambient conditions.		
\section{Implementation}
	\subsection{Analysis of the KPFM method}
	In KPFM, a signal, $S_{K}$, is generated by applying an AC voltage, $V_{AC}$, at frequency $f$ to a conductive tip above a grounded sample. A feedback loop applies a KPFM voltage, $V_{K}$, to the probe so that $S_{K}$ vanishes. The signal on which the KPFM feedback acts is:
		 \begin{align}
		  S_{K} = - \zeta_{j} (V_{K}+V_{0}),	
		  \label{eq:KPFM_feedback}
		 \end{align}
	where $V_{0} = V_{\textrm{tip}} - V_{\textrm{sample}}$ is the contact potential difference between the tip and sample when both are grounded, and $\zeta_{j}\equiv\zeta_{j}(V_{AC})$ is the sensitivity, which depends on the KPFM technique used (indicated through the subscript $j$), $V_{AC}$, the probe geometry, and imaging settings: such as the lift height (table of variables in Appendix \ref{sec:table_of_variables}). When $V_{K}=-V_{0}=V_{\textrm{sample}}-V_{\textrm{tip}}$, the signal vanishes. An image is created from the recorded $V_{K}$ as the cantilever raster scans the surface. The KPFM signal is written in the form of equation \ref{eq:KPFM_feedback} in order emphasize the similarity of H-KPFM to prior KPFM techniques and to facilitate their comparison. 
			
			\begin{figure}[tb]
				\centering
				\includegraphics[width=.45\textwidth]{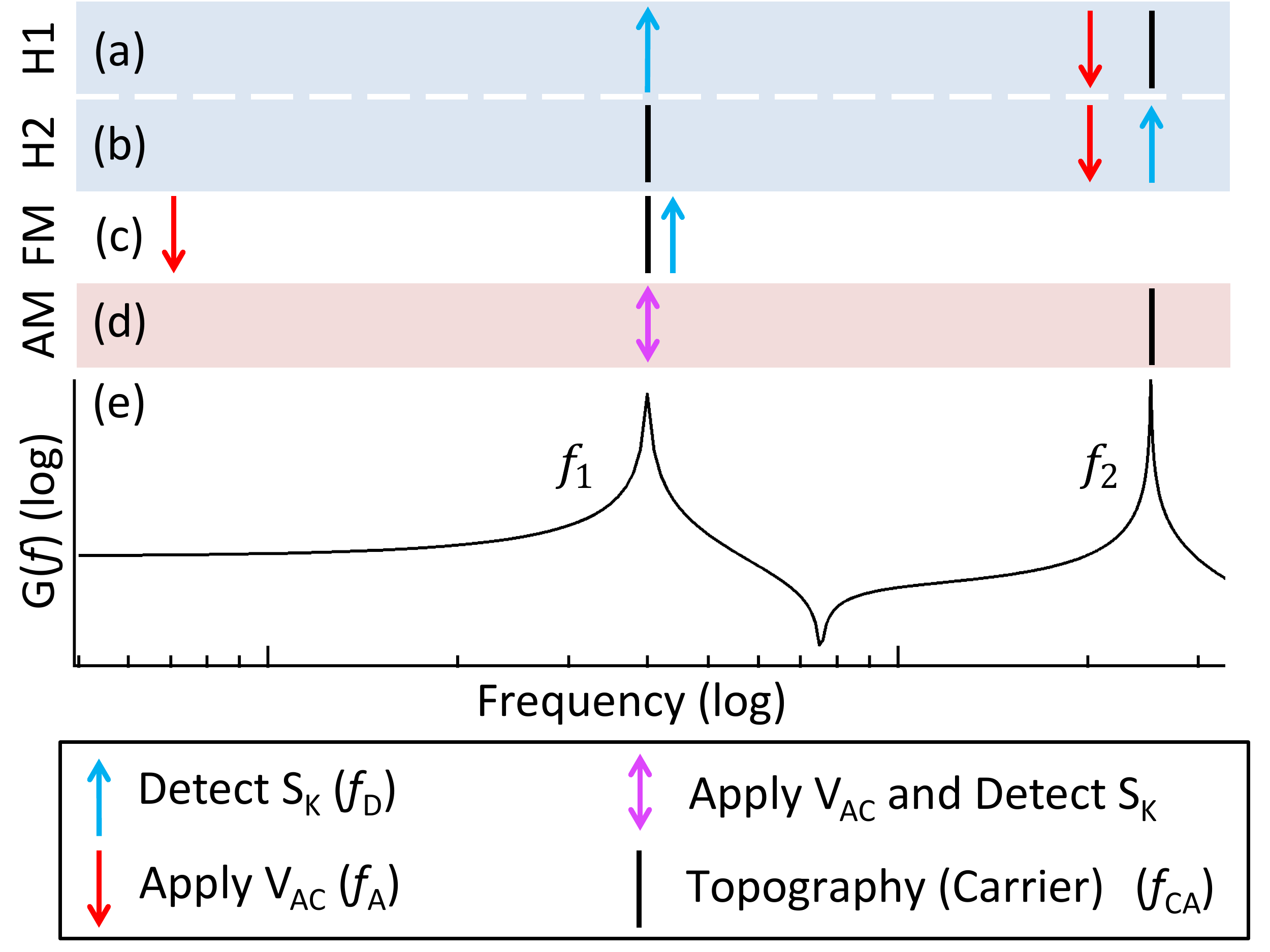}
				\caption{In H-KPFM (\textbf{\textcolor{light-blue}{$\blacksquare$}} a,b) an alternating voltage is applied at a frequency $f_{2}-f_{1}$  (\textbf{$\textcolor{red} \downarrow$}). The cantilever's response is {mixed with oscillation at the carrier frequency} in order to be detected at one resonance (${\textcolor{bright-blue} \uparrow}$). {The carrier oscillation occurs at} another resonance and is { also} used to maintain time-averaged distance to the surface ($\mid$). Likewise, in the sideband implementation of FM-KPFM ($\square$, c) a voltage is applied and the response detected at different frequencies: the alternating voltage is applied at $ f_{A} \ll f_{1}$ and detected at $f_{1}+f_{A}$. In AM-KPFM ({\textcolor{light-red}{$\blacksquare$}}), the alternating voltage is applied at the same frequency at which the cantilever response is detected ($\textcolor{purple}{\updownarrow}$). The magnitude of the cantilever transfer function $G(f)$ with each eigenmode modeled as a point-mass, is shown in (e).
				}
				\label{fig:CantileverResponse}
			\end{figure}

	In AM-KPFM, $S_{K}$ is detected at the same frequency as the applied $V_{AC}$ ($\textcolor{purple}{\updownarrow}$ in figure \ref{fig:CantileverResponse}), {\it{i.e.}} $f_{D}=f_{A}$ for AM-KPFM. Here we calculate the force above a conducting sample by modeling the tip-sample system as an metallic capacitor with energy $U = 1/2 C V^{2}$. The case for semiconductors is more complicated, but KPFM feedback operation is similar, and reduces to the metal case in the heavily-doped limit\cite{Hudlet1995}. The force on the cantilever has components at frequencies DC, $f_{D}$, and $2f_{D}$. The vertical force on the cantilever at frequency $f_{D}$ is then\cite{Jacobs1997}:
		\begin{align}
			F_{f_{D}} =  - C' V_{AC}(V_{K}+V_{0}),
			\label{eq:AM_force}
		\end{align}
	where $C' = \frac{dC}{dz}$.	We assume that the motion of each cantilever eigenmode is purely along the $z$-axis so that the transfer function of the cantilever $G(f)$ relates the driving force on the tip to the oscillation amplitude $A_{f}$ of the cantilever:
	\begin{align}
		A_{f_{D}} &= G(f_{D})F_{f_{D}}.
		\label{eq:amplitude}
	\end{align}
	 The optical lever sensitivity $\gamma(f)$ relates the signal generated at the photodetector to the amplitude of cantilever oscillation, so that:

	  	\begin{align}
	 	\label{eq:at_photodetector}
	 	S_{\text{photo}} &= \gamma(f_{D})A_{f_{D}}, \\
	 	&=  - \gamma(f_{D})G(f_{D})C' V_{AC}(V_{K}+V_{0}).\notag
	 	\end{align}

		 The signal from the photodetector is recorded by a quadrature lock-in amplifier with relative phase $\phi_{D}$:
	 \begin{align}
	 \label{eq:lock_in_signals}
	 S_{LIA}^{i}&= \text{Re}[- \gamma(f_{D})G(f_{D})C' V_{AC}(V_{K}+V_{0})e^{i\phi_{D}}],  \\
	 S_{LIA}^{q}&=  \text{Re}[- \gamma(f_{D})G(f_{D})C' V_{AC}(V_{K}+V_{0})e^{i(\phi_{D}+\frac{\pi}{2})}], \notag
	 \end{align}
	 where $S_{LIA}^{i}$ and $S_{LIA}^{q}$ are the in-phase and quadrature components of the signal, respectively, at the lock-in amplifier. The KPFM feedback loop operates on $S_{LIA}^{i}$, and when put in the form of equation \ref{eq:KPFM_feedback} is:
	\begin{align}
		  S_{K} &\equiv S_{LIA}^{i} = - \zeta_{AM}(V_{K}+V_{0}),
		  \label{eq:SKforAM}
	\end{align}
	where the sensitivity of AM-KPFM is:
	\begin{equation}
		\zeta_{AM} = \text{Re}[\gamma(f_{D}) G(f_{D}) C'V_{AC}e^{i\phi_{D}}].
		\label{eq:chi_am}
	\end{equation}	
	The relative phase of the lock-in amplifier, $\phi_{D}$, is adjusted in order to maximize the sensitivity for all techniques.

	\begin{table*}
		\caption{Example cantilever characteristics}
		\centering
		\begin{tabular}{ l | c  c  c  c  c  c  c  c }
			Name & $f_{1}$  (kHz)& $k_{1}$ (N/m) & \hspace{5 pt}$Q_{1}$\hspace{5 pt} &  $\gamma_{1}$ (V/nm)& \hspace{5 pt}$f_{2}$\hspace{5 pt}  & \hspace{5 pt}$k_{2}$\hspace{5 pt} & \hspace{5 pt}$Q_{2}$\hspace{5 pt} & \hspace{5 pt} $\gamma_{2}$\hspace{5 pt} \\
			\hline
			HQ:CSC35/Pt-C ($\mu$masch)& 130 & 5.0 & 230 & 0.030 & 810 & 88 & 440 & 0.070 \\
		\end{tabular}
		\label{table:cantilevers}
	\end{table*}
	In H-KPFM and FM-KPFM, the cantilever is shaken with amplitude {$A_{CA}$} at {the carrier} frequency {$f_{CA}$} by a non-electrostatic method (here, photothermally), $V_{AC}$ is applied at $f_{A}$, and the KPFM signal is detected at $f_{D}$ ($\mid$, \textbf{$\textcolor{red} \downarrow$}, and ${\textcolor{bright-blue} \uparrow}$, respectively, in figure \ref{fig:CantileverResponse}). The oscillation $A_{CA}$ is {used for topography control in single-pass mode, but} is {also} critical {for} the H-KPFM signal, {and} so  must be present, even when lift mode is used. We assume that the cantilever position is well-approximated by the sinusoidal motion at $f_{CA}$ (figure \ref{fig:CantileverResponse}), so that:
	\begin{align}
		(z-\bar{z}) =& A_{CA}\cos(2 \pi f_{CA}t + \phi_{CA}),
		\label{eq:expand_z}
	\end{align}
	where $z$ is the instantaneous tip-sample separation, $\bar{z}$ is the time-averaged separation, $A_{CA}$ is the amplitude of the {carrier}  oscillation, and $\phi_{CA}$ is the phase. Here we Taylor expand the tip-sample electrostatic force around its time-averaged height $\bar{z}$ so that the capacitive force on the cantilever is\cite{Sugawara2012}:
		\begin{align}
			F = \Bigg[ C'+C''A_{CA}\cos(2 \pi f_{CA}t + \phi_{CA})\Bigg] \notag \\
			\times \frac{[V_{AC}\cos(2\pi f_{A}t +\phi_{A})+V_{K}+V_{0}]^{2}}{2}.
			\label{eq:All_u_terms}
		\end{align}

	As in AM-KPFM, a term linear in $V_{AC}$ is used for the KPFM feedback, and there are three frequencies at which such a signal is generated: $f_{A}$, $f_{A}+f_{CA}$, and $|f_{A}-f_{CA}|$. The force at the first frequency, proportional to $C'$, is used for AM-KPFM (see equation \ref{eq:AM_force}), while the forces at the second and third frequencies, each proportional to $C''$, are used for H-KPFM. Then, up to a phase shift, each force is:
	\begin{align}
	F_{f_{D}} = -\frac{C^{''}A_{CA}}{2}V_{AC}(V_{K}+V_{0}).
	\label{eq:H_KPFM}
	\end{align}
	Like Sugawara {\it et al}. \cite{Sugawara2012}, we choose $f_{D} = f_{A}+f_{CA}$. The case $f_{D} = |f_{A}-f_{CA}|$ results in an equivalent force. Then, as with AM-KPFM above, the signal used for H-KPFM feedback depends on the cantilever transfer function and the optical lever sensitivity at the detection frequency, so that the signal at the photodiode is, up to a phase shift:
	\begin{align}
	S_{\text{photo}} &= -\frac{\gamma(f_{D})G(f_{D})C^{''}A_{CA}}{2}V_{AC}(V_{K}+V_{0}).
	\label{eq:signal_H}
	\end{align}
	
	Once the phase shift is included, the H-KPFM feedback signal is put in the form of equation \ref{eq:KPFM_feedback} with sensitivity:
	\begin{align}
	\zeta_{H}=\text{Re}\biggl[\frac{\gamma(f_{D})G(f_{D})A_{CA}}{2}C^{''}V_{AC}e^{i(\phi_{CA}+\phi_{A}+\phi_{D})}\biggr].
	\label{eq:heterodyne_chi}
	\end{align}
	Thus the sensitivity of H-KPFM differs from AM-KPFM both by it dependence on $C''$ instead of $C'$ and by its dependence on the { carrier} oscillation amplitude $A_{CA}$.  If it is necessary to scan far from the surface, $A_{CA}$ can be increased to enhance sensitivity. Note that FM-KPFM similarly depends on $A_{CA}$\cite{Zerweck2005}.

	In H-KPFM, both the detection frequency, $f_{D}$ and the {carrier oscillation} frequency, $f_{CA}$, are free to be chosen, and once chosen, determine the frequency at which $V_{AC}$ is applied, $f_{A}$.
	Earlier works on H-KPFM considered the case $f_{CA} = f_{1}$, the first cantilever resonance, and $f_{D} = f_{2}$, the second cantilever resonance\cite{Sugawara2012,Ma2013,Ma2013a}. In this article, this implementation is called "H2" for heterodyne amplified by the second cantilever resonance. Here the case $f_{CA} = f_{2}$, $f_{D} = f_{1}$ is also considered, for enhanced sensitivity, and we call it "H1" because $S_{K}$ is amplified by the first resonance. 
	
\subsection{Experimental Setup}

	All methods are implemented on a commercial AFM (Cypher, Asylum Research). The motion of a platinum-coated cantilever is measured with an optical lever employing a 860 nm laser and detected by a quad-photodiode.  The optical lever sensitivity is determined for each eigenmode from amplitude vs. distance curves, and the spring constants are determined by fitting the cantilever's thermal spectrum (table \ref{table:cantilevers}).
	
	\begin{figure}[!htb]
				\centering
				\includegraphics[width=.45\textwidth]{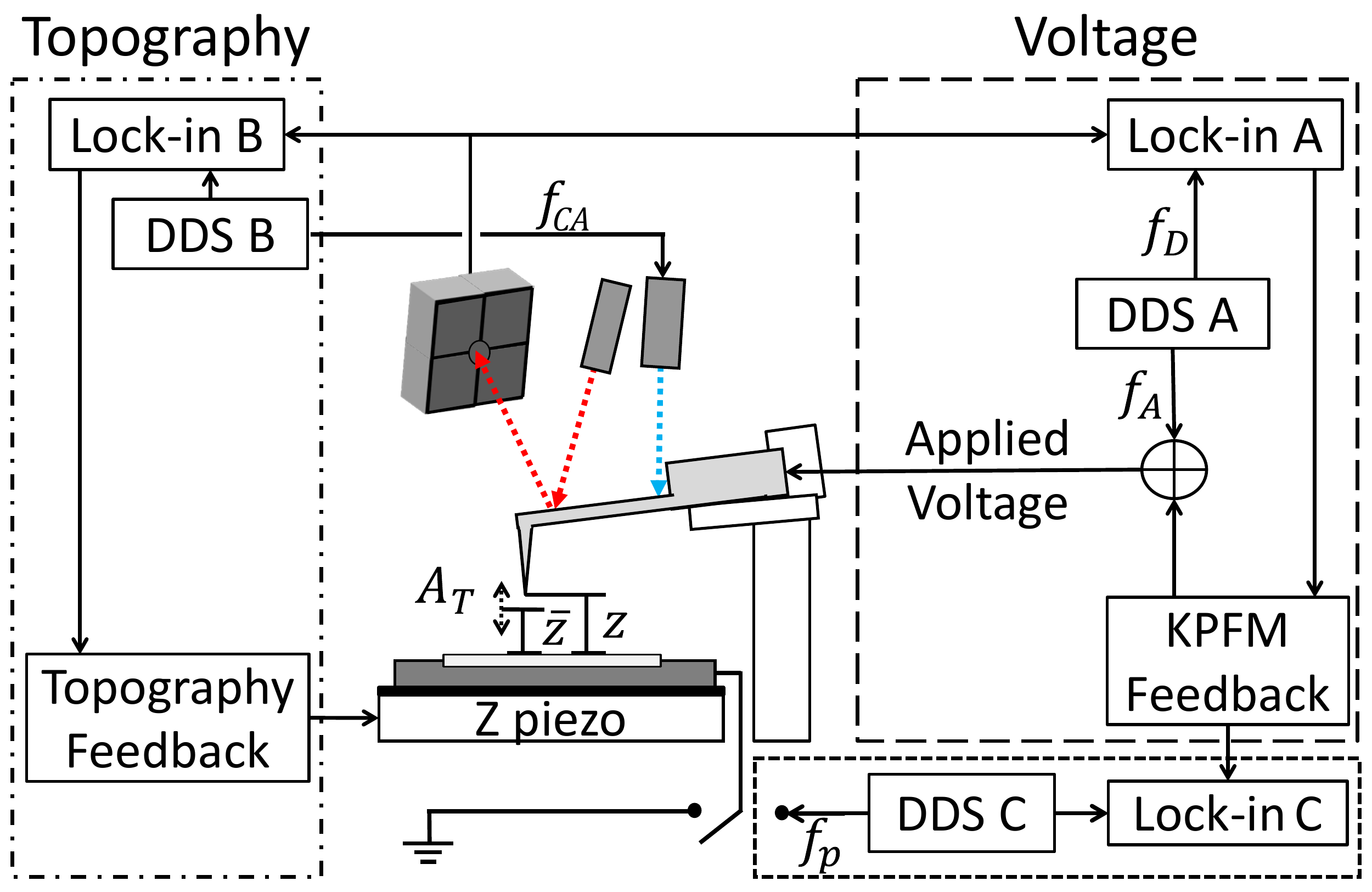}
				\caption{A feedback loop controls the separation between a photothermally driven cantilever and the sample through the cantilever's oscillation amplitude by adjusting the sample height (left). A voltage $V_{AC}$ at frequency $f_{A}$ is added to $V_{K}$, the KPFM voltage, and applied to the probe. The cantilever oscillation at $f_{D}$ is then detected by lock-in amplifier B and used by a feedback loop to control the DC voltage applied to the cantilever (right). A third lock-in amplifier measures the response of $V_{K}$ to a perturbation in order to deduce the KPFM transfer function.	
				}
				\label{fig:HAMKPFMsetup}
	\end{figure}
	
	KPFM is implemented using two direct digital synthesizers (DDS), each paired with a lock-in amplifier (LIA). In particular, the cantilever is excited at $f_{CA}$ photothermally by DDS B (figure \ref{fig:HAMKPFMsetup}) for topography control. DDS A generates an AC voltage at frequency $f_{A}$ that is applied to the probe. LIA A detects the cantilever's oscillation at $f_{D}$. The relative phases of signals from DDS A and B are maintained through the synchronization of the AFM's internal clock. To measure the transfer function of the KPFM loop, DDS C is used to apply an AC voltage, $V_{p}$ to the substrate. LIA C detects the response of $V_{K}$ to the perturbation. 
	
	AM feedback is used for the topographical loop for all KPFM methods. In our earlier experiment\cite{Garrett2015}, an FM feedback loop controlled the tip-substrate distance while maintaining attractive-mode scanning\cite{Garcia1999}. Although FM topography feedback is adapted from the original implementation of H-KPFM\cite{Sugawara2012}, it contains one major disadvantage: the frequency shift is a non-monotonic function of distance\cite{Giessibl2003} and so the tip collides with the surface when its motion deviates too far from the topography setpoint. With AM topography control the feedback operates on a signal that is monotonic with distance, except at one bistability that can be avoided\cite{Garcia1999}. When AM feedback is used for topography, small perturbations, that once destroyed probes, no longer affect scan stability.
			
	The settings for the different KPFM techniques are chosen to realistically represent each technique's capabilities and are similar to those of previous experiments\cite{Glatzel2003,Zerweck2005}. FM-KPFM is implemented with sideband detection\cite{Zerweck2005}: $f_{CA} = f_{1}$ and $f_{D} = f_{1}+ f_{A}$, and the modulation frequency $f_{A} = 2$ kHz is maintained. For AM-KPFM, $V_{AC}$ is applied at $f_{1}$, and the topography loop operates at $f_{2}$. For H-KPFM, the H1 implementation uses $f_{CA} = f_{2}$ and $f_{D} = f_{1}$, while $f_{CA} = f_{1}$ and $f_{D} = f_{2}$ for H2 (see figure \ref{fig:CantileverResponse}). For all methods, $V_{AC}=1 \text{ V}.$ 
			
	All scans are performed on a micron-sized flake of few-layer graphene (FLG) on boron doped silicon with a thin native oxide layer (15-25 Ohm-cm, Virginia Semiconductor), prepared by exfoliation\cite{Novoselov2004}. Both flakes of Highly Ordered Pyrolytic Graphite (HOPG) and FLG are observed with AFM. The HOPG is a few tens of nm tall and causes band bending in the Si surface potential at its edges but has negligible patch potentials. The FLG is $\approx$ 1 nm high and does not change the surface potential of Si around it but is covered with patch potentials. Because the FLG/Si boundary has less topography change, and a surface potential profile that is symmetric around the boundary, it is chosen for the following measurements.
	\subsection{Eliminating artifacts}
\begin{figure}[!htb]
		\centering
		\includegraphics[width=.45\textwidth]{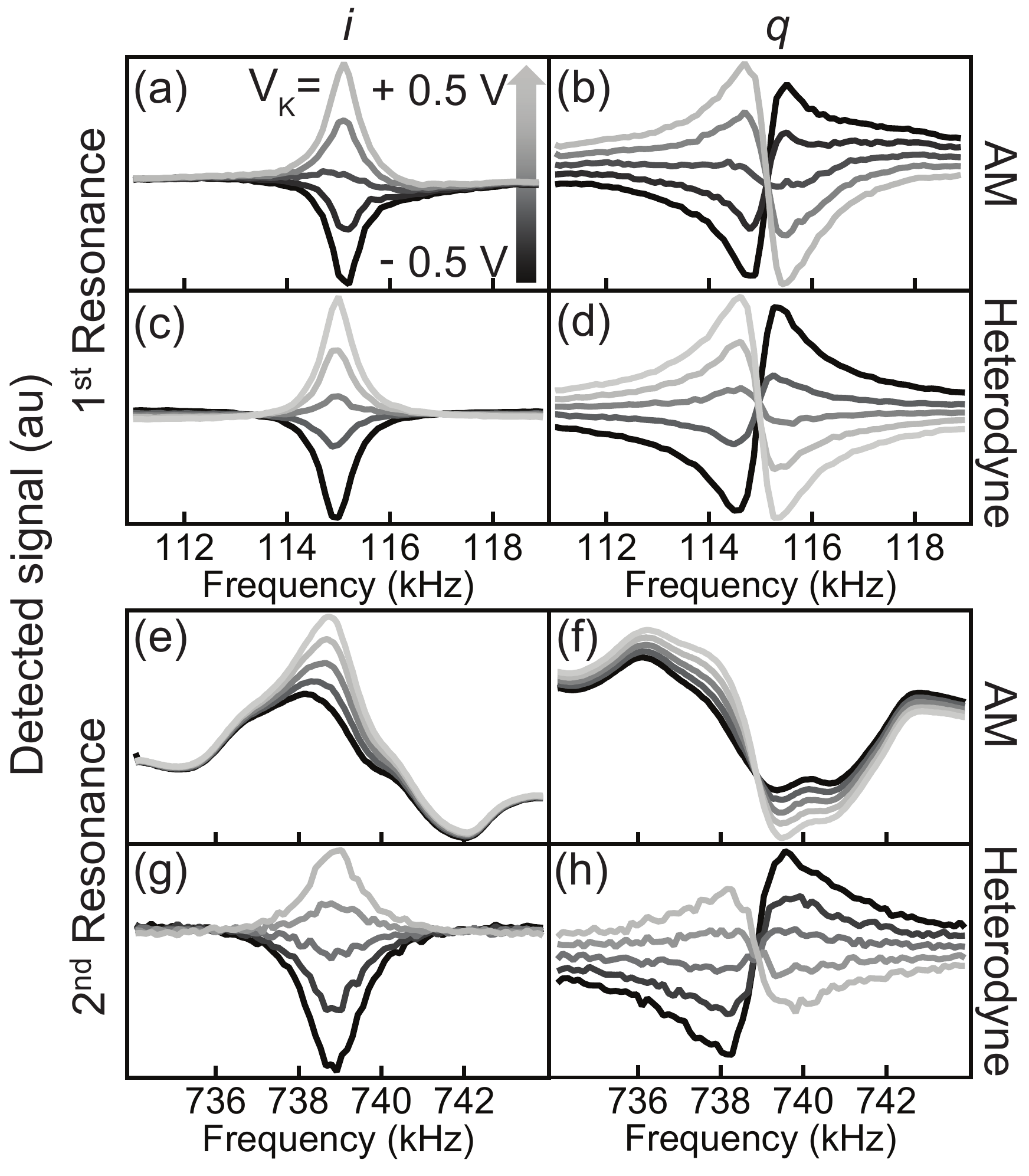}
		\caption{H-KPFM removes the distortion caused by AC coupling in the KPFM signal vs. detection frequency curves ($S^{i}_{LIA}$ vs $f_{D}$ and  $S^{q}_{LIA}$ vs $f_{D}$). (a,b) At the first eigenmode, the in-phase ($i$) and quadrature ($q$) components of the AM-KPFM signal show little distortion, but (e,f) at the second eigenmode, distortion due to inductive AC coupling between the KPFM voltage and tip holder, which increases with frequency, is observed. (c,d,g,h) Heterodyne excitation generates no distortion. The KPFM voltage applied to the tip, $V_{K}$, is sampled at values above and below the contact potential difference. Similar measurements were used to detect AC coupling in \cite{Diesinger2008}.}
		\label{fig:FreqCurves}
	\end{figure}

	\begin{table*}
		\small
		\caption{Common artifacts in Kelvin Probe Force Microscopy}
		\centering
		\begin{tabular}{ l | l |  c | c | c }
			\hline
			Type & Example Source & H & FM & AM \\
		
			\hline
			Extraneous Signal ($S_{E}$)&&&&\\
			\hspace{10 pt}\textit{Time-independent}  & AC inductive coupling, between $V_{AC}$ and piezo (figure 				\ref{fig:FreqCurves})\cite{Barbet2014}&-&-& $\times$\\
		
			\hspace{10 pt}\textit{Periodic}&  Topographical oscillation detected in voltage bandwidth (figure \ref{fig:V_speed}i)& - & $\times$ & -\\
			
			\hspace{10 pt}\textit{Intermittent} & Collision with surface &$\times$ &$\times$&$\times$\\
			\hline 
			Stray Capacitance &  Long-range electrostatic force from cantilever \cite{Jacobs1997,Colchero2001,Ma2013}&-&-& $\times$\\ 
			\hline
		
		\end{tabular}
	
		\begin{tabular}{c}
			Legend:	$\times$ =  large artifact, - = small artifact 
		\end{tabular}
		\vspace*{-\baselineskip}
		\label{table:artifacts}
	\end{table*}
	Several artifacts originate from signals interfering with the Kelvin probe signal, $S_{K}$ \cite{Kalinin2000,Diesinger2008,Melin2011,Barbet2014}. Examples of such signals include AC coupling between $V_{AC}$ and a piezo in the cantilever holder (figure \ref{fig:FreqCurves}) or detection of the topography oscillation (at $f_{CA}$) within the lock-in amplifier (LIA) bandwidth (table \ref{table:artifacts}). The resulting signal detected at the LIA contains both the desired signal, $S_{K}$, and an extraneous signal, $S_{E}$, and is given by:
	\begin{align}
	S_{LIA}^{i} = S_{K}+ S_{E}.
	\label{eq:Extra_signal}
	\end{align}
	A setpoint, $S_{sp}$ for the voltage feedback loop is chosen to compensate for $S_{E}$ (above we assume $S_{E}=0$, and so a setpoint is not needed). When both $S_{E}$ and $S_{sp}$ are included, the Kelvin probe loop detects the voltage:
	\begin{align}
	V_{K} = -V_{0} + V_{E},
	\label{eq:extra_voltage_inVK}
	\end{align}
	which contains an extraneous voltage:
	\begin{align}
	V_{E} =  \frac{S_{E}-S_{sp}}{\zeta_{j}}.
	\label{eq:V_error}
	\end{align}
	The topography is imprinted on $V_{K}$ through the height-dependence of $\zeta_{j}$, the sensitivity from equation \ref{eq:KPFM_feedback}, which complicates attempts to remove the artifact in post-processing\cite{Barbet2014}. 
		
	\begin{figure}[!ht]
		\centering
		\includegraphics[width=.45\textwidth]{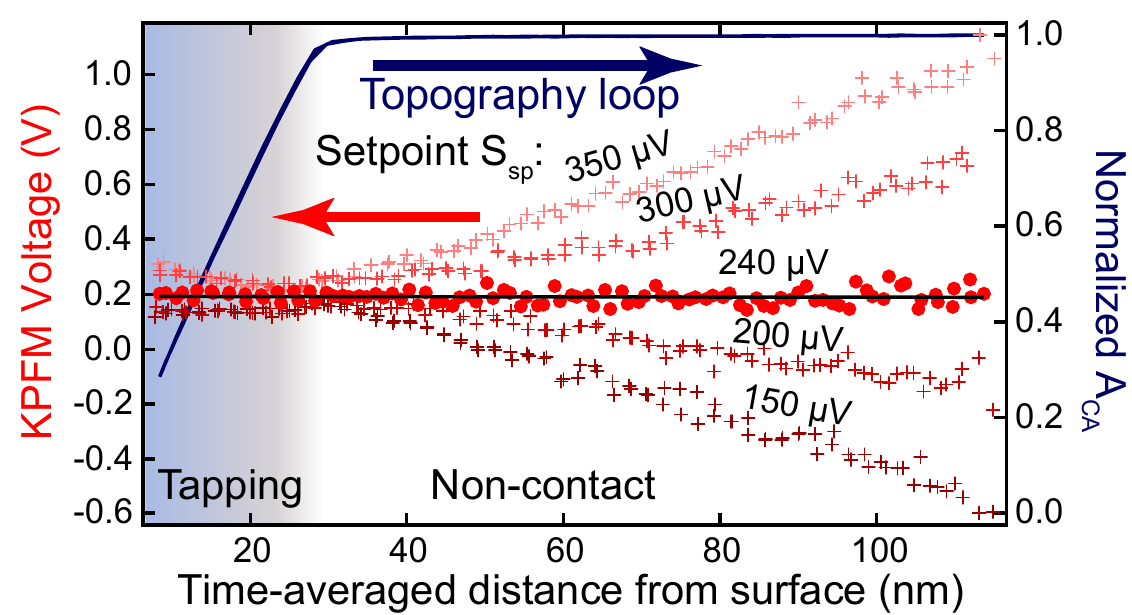}
		\caption{The setpoint of the KPFM feedback loop is adjusted to minimize the height-dependence of the KPFM voltage, $V_{K}$. Here the setpoint is swept from 150 to 350 $\mu$V over several approaches. The center curve (240 $\mu$V, $\color{red}\bullet$) shows a distance dependence of only 0.03 mV/nm. The dark blue curve indicates the topographical oscillation amplitude normalized to its value far from the surface. }
		\label{fig:Distance_curves}
	\end{figure}
		
		Conversely, the height dependence of $V_{E}$ can also be used to identify $S_{E}$. If $S_{E}$ is small enough and does not vary in time, $S_{sp}$ can be chosen so that the numerator of equation \ref{eq:V_error} vanishes.
		In this paper, the height dependence of $V_{K}$ is used to choose $S_{sp}$. If a sample has uniform surface potential, then:
		
		\begin{align}
		\frac{dV_{K}}{d\bar{z}} = \frac{dV_{E}}{d\bar{z}} = \frac{S_{sp}-S_{E}}{\zeta_{j}^{2}}\frac{d\zeta_{j}}{d\bar{z}}.
		\label{eq:calibratin}
		\end{align}
		If $dV_{K}/d\bar{z}\approx 0$, then $S_{sp}\approx S_{E}$, as $d\zeta_{j}/d\bar{z}$ does not vanish. 
		
		To minimize $V_{E}$, the KPFM feedback setpoint, $S_{sp}$, is varied over a range of 200 $\mu$V, and a $V_{K}$ vs. height curve is recorded for each $S_{sp}$ (figure \ref{fig:Distance_curves}). For most $S_{sp}$, the measured $V_{K}$ does depend on height, indicating that $S_{sp}\neq S_{E}$. The variation amongst the curves decreases when the tip-sample separation is reduced (until intermittent contact with the sample begins at $\approx$ 20 nm). The setpoint with the least distance dependence (240 $\mu$V), is maintained for the KPFM scans. The offset originates at the output of the low-pass filter on the lock-in amplifier for our setup, and it varies slightly from day to day, so the calibration must be repeated for every set of measurements.

\section{Resolutions}\label{sec:on-resolution}

The temporal, voltage, and spatial resolutions of the different KPFM implementations are compared through several tests, the results of which are summarized in table \ref{table:resolutions}. 

\subsection{Time Resolution}

H-KPFM achieves fast time resolution by avoiding several artifacts that limit speed of the other KPFM techniques (Table \ref{table:artifacts}). {Because several limits on KPFM time resolution are proportional to $f_{D}$, such as the bandwidth of a cantilever resonance ($f_{D}/2Q$) and the Nyquist frequency ($f_{D}/2$),} higher resonant frequencies are expected to increase bandwidth{. However}, for AM-KPFM, {higher frequencies} also increase the AC coupling\cite{Barbet2014} (figure \ref{fig:FreqCurves}).  AC coupling does not affect H-KPFM or FM-KPFM as significantly because the applied and detected signals are at different frequencies. Consequently, H-KPFM can employ cantilevers with higher resonant frequencies than AM-KPFM. This limitation of AM-KPFM is due to the drive piezo that is present in most cantilever holders. Additional circuitry can mitigate this artifact\cite{Diesinger2008,Melin2011,Polak2014}, but typically the circuitry must be custom-made.

On the other hand, the artifact that limits FM-KPFM scan speed is fundamental to its operation. In both H- and FM-KPFM, carrier and KPFM signals must be present at the same time. If $A_{CA}=0$, then $S_{K}$ vanishes, even in lift mode (equation \ref{eq:signal_H}, \cite{Zerweck2005}). FM-KPFM scan speed is limited by a periodic $S_{E}$ imprinted on the KPFM signal because the two signals, at $f_{D}$ and $f_{CA}$, are so close in frequency space. Then the extraneous signal is estimated by considering how the cantilever oscillation $A_{CA}$ at $f_{CA}$ is detected by a lock-in amplifier with reference signal at $f_{D}$. When the signal is input into equation \ref{eq:V_error}, the extraneous voltage is:
\begin{align}
V_{E}^{CA} = \frac{\gamma(f_{CA})A_{CA}}{\zeta_{j}} \text{Re}\Bigg[\frac{e^{i(2\pi f_{A} t)}}{1+if_{A}/B}\Bigg],
\label{eq:FM_error}
\end{align}
where $B$ is the bandwidth of the LIA's low-pass filter, and we set $S_{sp}=0$ for simplicity. In typical KPFM operation, the prefactor, $\frac{\gamma(f_{CA})A_{CA}}{\zeta_{j}}$, is large compared to the surface voltage contrast being measured. To reduce $V_{E}^{CA}$ then, $B$ must be chosen so that $B\ll f_{A}$. For H-KPFM $f_{A}>$100 kHz, so the bound on $B$ is large. FM-KPFM, however, typically works with $f_{A}\approx1-3$ kHz, which limits $B$. $V_{E}^{CA}$ decreases with increasing $f_{A}$, which can be used to increase the available bandwidth even though it concurrently decreases the sensitivity { because $|G(f_{1}+f_{A})|$, which is proportional to the sensitivity, decreases with increasing $f_{A}$}. Note also that $V_{E}^{CA}$ is periodic in time, and so it cannot be mitigated by varying the KPFM feedback loop setpoint. 

Previous measurements of time resolution either investigate the KPFM feedback loop response to a periodic voltage applied to the setpoint\cite{Zerweck2005}, or substrate \cite{Diesinger2010}, or how quickly a well-characterized sample can be scanned while retaining KPFM contrast\cite{Sinensky2007}. Here the former method is used to estimate the cut-off frequency, $f_{c}$, which is defined as the frequency at which the KPFM loop response has dropped to $\approx 71\%$ of the low-frequency response (-3 dB). In table \ref{table:resolutions}, the cut-off time, $t_{c} =  1/f_{c}$, is listed instead, so that smaller values indicate a better resolution.
\begin{figure}[!ht]				
	\centering							
	\includegraphics[width=.45\textwidth]{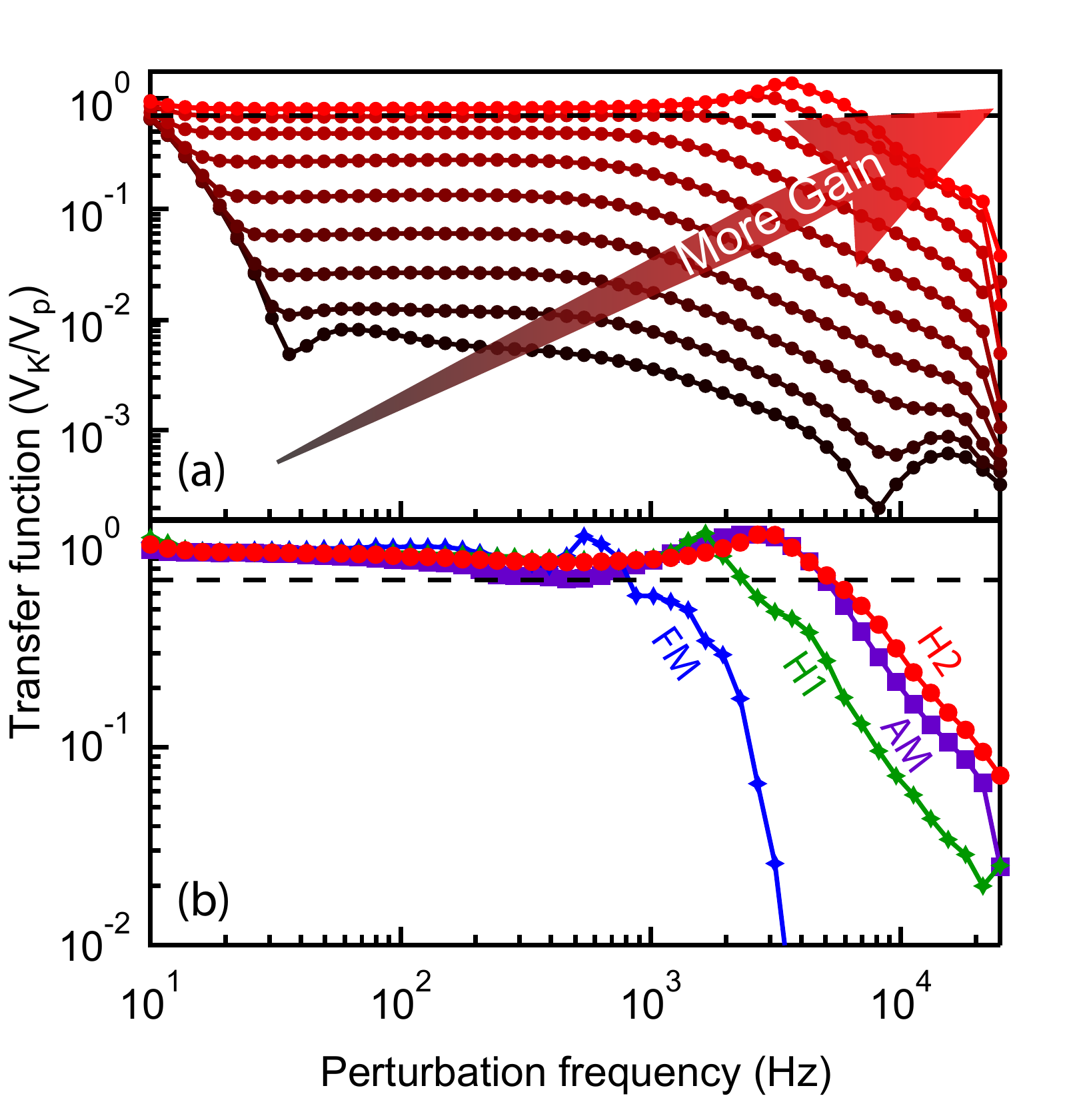}
	\caption{ The transfer functions for the different KPFM methods are measured by applying a periodic voltage perturbation to the substrate and recording the response of the KPFM loop. (a) The transfer function of a H-KPFM voltage feedback loop increases and becomes more uniform as the gain is increased (H2 implementation). The proportional gain is increased 3 orders of magnitude, and overshoot is constrained to $\frac{V_{K}}{V_{p}}\leq1.2$. A dashed line indicates the -3 db point used to calculate the cut-off frequency, $f_{c}$. (b) The cut-off frequency depends strongly on the method used, varying by almost an order of magnitude. The gain for each of the methods is chosen by the same optimization procedure.
	}		
	\label{fig:transfers}
	
\end{figure}

The reported time resolutions of AM-KPFM typically exceed those of FM-KPFM, even though the specific resolution depends on both the cantilever and atomic force microscope used. Of the references discussed here, a few optimize temporal resolution for their AFMs\cite{Sinensky2007,Diesinger2010}. For the others, the speeds cited are typical of an imaging method rather than the outcome of an optimization procedure. Diesinger {\it et al.}\cite{Diesinger2010} report an implementation of AM-KPFM that achieves $f_{c}\approx200$ Hz, limited by the analog-digital conversion of the KPFM loop. In air, Sinensky and Belcher demonstrate that AM-KPFM can maintain some voltage contrast at scan speeds up to $1,172 \text{ }\mu \text{m s}^{-1}$, by scanning 2-$\mu$m wide stripes of DNA\cite{Sinensky2007}. In the language used here, that corresponds to $f_{c}\approx1.2$ kHz. FM-KPFM is reported to operate with similar speed when either in the sideband ($f_{c}\approx35 $ Hz) or phase locked loop ($f_{c}\approx30$ Hz) is used, even though the sources of speed limitation are dissimilar\cite{Zerweck2005,Diesinger2014}. Recent improvements to the KPFM feedback increase the cut-off frequency to 100 Hz with a larger modulation frequency (4 kHz)\cite{Wagner2015a}. Reported open-loop FM-KPFM scan speeds include $0.85$ $\mu$m$s^{-1}$ (or 5 min per (500 nm)$^{2}$, 256$\times$256 pixel scan, trace and retrace)\cite{Borgani2014a} and $1.3$ $\mu$m$s^{-1}$(or 3 min per (450 nm)$^{2}$, 256$\times$256 pixel scan, trace and retrace)\cite{Takeuchi2007}. 

To measure the closed loop transfer function of each KPFM method, an AC voltage ($V_{p}=1$ V at perturbation frequency $f_{p}$) is applied to the substrate by a third DDS, while the cantilever height is maintained at the surface by a topographical feedback loop, with $A_{CA}\approx8$ nm. The KPFM loop tracks the voltage, and $V_{K}(f_{p})$ is detected by a third lock-in amplifier. The frequency is swept from $f_{p}$ = 10 Hz to 25 kHz. The proportional gain of the control loop is increased until the bandwidth stops increasing, and the integral gain is then increased until the transfer function is flat across its bandwidth (figure \ref{fig:transfers}). The cutoff frequencies for H2, H1, and AM are $5.3, 2.3,$ and $5.0$ kHz, respectively (table \ref{table:resolutions}). By further optimizing the feedback loops the bandwidth might be increased\cite{Diesinger2010,Bechhoefer2005,Wagner2015a}.

The measurement of the FM-KPFM transfer function is complicated by the presence of the topological feedback signal near the KPFM signal, which causes $V_{K}$ to include an extraneous, rapidly oscillating voltage (see equation \ref{eq:FM_error}). The separation between the KPFM signal and the interfering topography signal is equal to the $f_{A}$ of FM-KPFM, which here is 2 kHz, quite typical for FM-KPFM\cite{Glatzel2003,Zerweck2005}. First the transfer function is measured with only the lock-in amplifier's own low-pass filter, but the extraneous voltage is so large that it overwhelms the signal until the frequency of the low-pass filter is decreased to 700 Hz, giving $f_{c}\approx400$ Hz. However, the extraneous voltage imprinted by the topography signal remains $\approx400$ mV, prohibitively large for practical measurements. Second, a notch filter is placed on the lock-in amplifier at $f_{A}$ (2 kHz) in order to further mitigate $V_{E}^{CA}$. The notch filter both decreases $V_{E}^{CA}$, and also allows the filter on the lock-in amplifier to be increased to 1 kHz. In this configuration the cutoff frequency of FM-KFPM is determined to be $f_{c}\approx820$ Hz. {It is worth noting that the measured bandwidths appear to exceed the bandwidth of the cantilever resonances, $f_{i}/2Q_{i}$. It is possible that the feedback of closed-loop methods flattens the transfer function analogously to the way the transfer function of an operational amplifier is flattened by placing a resistor across it\cite{Bechhoefer2005}; however, further investigations are warranted.}

\begin{figure}[!ht]
		\centering
		\includegraphics[width=.45\textwidth]{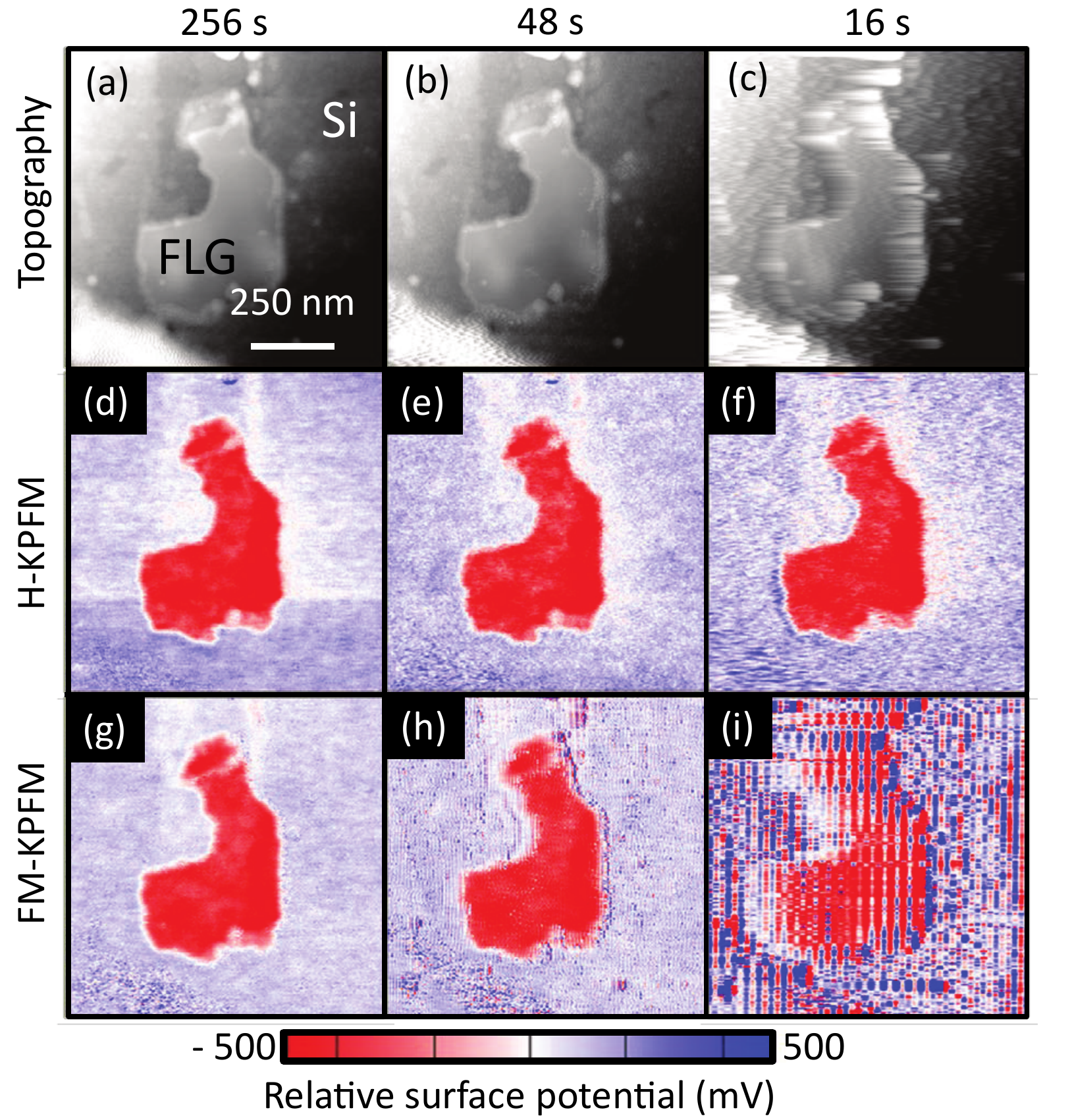}
		\caption{Images show the affect of increasing scan speed. (a-c) Topography images of the few layer graphene (FLG) become blurred. (d-f) H-KPFM allows the potential of FLG to be imaged with increasing speed and minimal distortion. (g-i) The topography oscillation is imprinted on the potential image when the bandwidth is increased if FM-KPFM is used. Topography scanning may be the primary speed limit. The height of the FLG is  $\approx2$ nm. All scans are 256$\times$256 pixels.}
		\label{fig:V_speed}
	\end{figure}

To investigate how $f_{c}$ translates into imaging speed, a few-layer graphene (FLG) flake is scanned with H- and FM-KPFM while the line scan speed is increased from 1 Hz to 79 Hz, over a 1$\times$1 $\mu$m area with 256$\times$256 pixels with $A_{CA}$ = 16 nm (figure \ref{fig:V_speed}). By 4 Hz (48 s per frame), FM-KPFM shows stripes. { To investigate the cause of these stripes, the FLG is imaged without the aforementioned notch filter at 2 kHz. At 8 Hz, the amplitude of the stripes is $< 0.3$ V with the notch filter, but rises to $> 1.5$ V when the notch filter is removed. Thus the signal $V_{E}^{CA}$ does contribute to the stripe artifact, although the details of the feedback loop likely influence the stripes as well.} At higher frequencies, the FM feedback loop oscillates wildly near the edges of the FLG.

With H-KPFM, on the other hand, clear contrast is maintained up to 16 Hz (16 s per frame), and at higher frequencies some contrast is maintained. However, the topographical feedback loop stops tracking the surface, and topographical inconsistency affects the potential image. At 79 Hz, patches on the graphene flake are no longer visible. A similar limitation due to topographical feedback loop speed is reported in \cite{Sinensky2007}. 

\begin{table*}[ht]
	\small
	\caption{Resolutions that characterize KPFM}
	\centering
	\begin{tabular}{l | c  | l  | r  r  r  r r}
		\hline
		Resolution & Figure of Merit  & Definition  & \hspace{10 pt} H2    & \hspace{10 pt}H1    & \hspace{10 pt}FM    & \hspace{10 pt}   AM & (units)\\
		\hline
		Time* &  $t_{c}=1/f_{c}$  & Closed-loop 3 dB cut-off time\cite{Diesinger2010} & 0.19 & 0.43 & 1.2 & 0.20 & ms\\
		&&&&&&&\\
		\hline
		Voltage** & $V_{m}$ &  $V = V_{K}+V_{0}$ for which signal = noise\cite{Nonnenmacher1991}& 73 & 41 & 96 & 2.0& mV\\
		& &   & &&&&\\
		\hline
		Space** & $l_{10-90}$ & Distance from boundary over which voltage & 45 & 42 & 49 & 68& nm\\ & & changes from 10\% to 90\% \cite{Zerweck2005} &   &&&&\\
		\hline
	\end{tabular}
	
	\begin{tabular}{c}
		*At $A_{CA}=8$ nm, $V_{AC}=1$, $V_{p}=1$ V, and at the surface, with topographical feedback on\\
		**At $A_{CA}=8$ nm, Bandwidth = 200 Hz, $V_{AC}$=1 V, and lift height 11 nm\\
	\end{tabular}
	\label{table:resolutions}
\end{table*}
\subsection{Voltage Resolution}	

\subsubsection{Accuracy}
\label{sec:Accuracy}

 Whereas the tip apex detects the potential directly beneath it, the inclusion of stray capacitance from the cantilever results in surface potential spatially averaged over many microns (about the width of the cantilever)\cite{Jacobs1997,Koley2001,Colchero2001,Ma2013}. The unknown and varying relative capacitances of the tip apex and cantilever limit AM-KPFM to qualitative contrast in most conditions\cite{Jacobs1997,Colchero2001}. Both H-KPFM and FM-KFPM mitigate the stray capacitance effect through their dependence on $C''$ rather than $C'$\cite{Glatzel2003,Ma2013}. Here the stray capacitance must be assessed in order to understand the relation between the measured potential sensitivity and the ability to actually distinguish between two nanoscale objects. 
\begin{figure}[h]				
	\centering							
	\includegraphics[width=.45\textwidth]{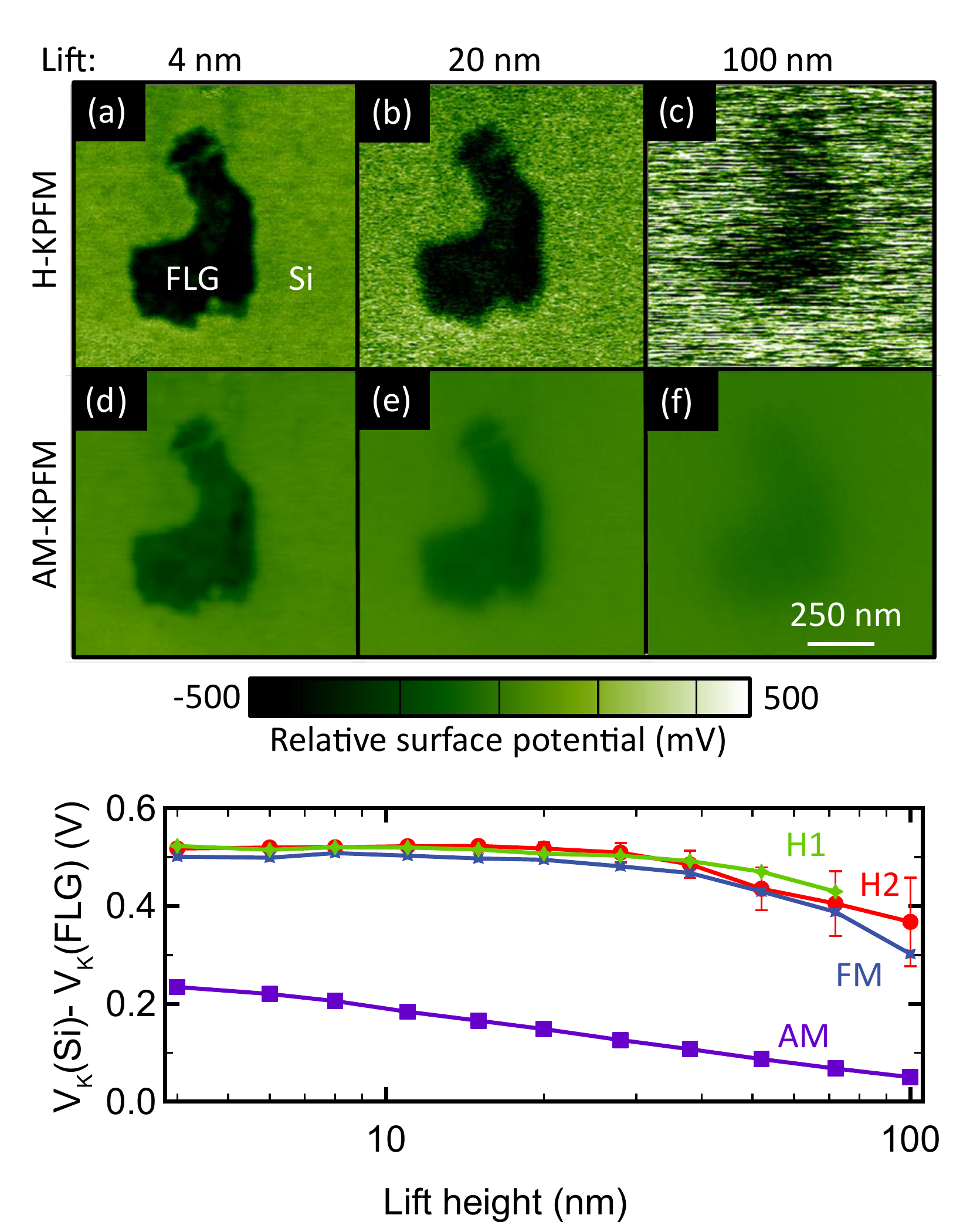}
	\caption{The voltage contrast between a few layer graphene (FLG) flake and Si substrate reveals the stray capacitance effect. (a-c) The voltage contrast between (few-layer) graphene and silicon changes little as the probe height increases from 4 to 100 nm for H-KPFM. (d-f) However, for AM-KPFM, the contrast at 100 nm differs by a factor of five from that at 4 nm. (g) A comparison between the four different KPFM methods shows that methods that depend on $C''$ more accurately represent the potential contrast than AM-KPFM, which depends on $C'$.
	}		
	\label{fig:Distance_and_dv}
\end{figure}
The capacitance of tip and cantilever changes with tip-sample separation, and consequently, so does the measured average voltage contrast between the FLG flake and Si substrate,  $\Delta V$. The correspondence between the actual and measured potentials is tested by observing the change of $\Delta V$ with lift height, akin to \cite{Spadafora2011}. At closest approach the tip apex capacitance dominates. As the tip-sample separation is increased, $\Delta V$ changes little until the proportion of capacitance due to the apex decreases to a value comparable to the cantilever capacitance contribution. The criterion of $\frac{d\Delta V}{d\bar{z}}\approx0$ near the surface is adopted to ensure that the apex contribution dominates. In the limit of large lift height, the cantilever contribution dominates, and no potential contrast is observed. The potential contrast is estimated for each height by calculating the difference between the average potential inside the FLG/silicon boundary (figure 11e) and the average potential outside.

The contrast between Si and FLG changes little for H- and FM-KPFM, as the cantilever lift height is varied (figure \ref{fig:Distance_and_dv}a-c,g). On the other hand, the AM-KPFM detected voltage contrast changes by a factor of five as the lift height is decreased from 100 to 4 nm (figure \ref{fig:Distance_and_dv}d-f,g). Thus the average potential contrast measured with $C''$ methods is more accurate than the contrast measured by AM-KPFM.

\subsubsection{Minimum detectable voltage}

The minimum detectable voltage, $V_{m}$, is the tip-sample voltage difference at which the signal is equal to the noise\cite{Nonnenmacher1991,Giessibl2003,Sugawara2012,Ma2013a}. Here $N(B)$ is the noise power in the signal $S^{i}_{LIA}$ within the bandwidth $B$. The minimum detectable voltage for any KPFM method is:
\begin{align}
	V_{m} =\frac{\sqrt{N(B)}}{\zeta_{j}}.
	\label{eq:Vm}
\end{align}
Note that $N(B)$ increases as the bandwidth increases. Thus increasing temporal resolution restricts voltage resolution.

The sources of noise in an AFM can be divided into three categories\cite{Labuda2012b}. The first, detection noise, includes angular fluctuations of the light beam and optical shot noise. The second, displacement noise, includes the reaction of the topography feedback loop to perturbations, such as 60 Hz line noise or the voltages applied in KPFM. The third, force noise, includes Brownian motion and stresses caused by light optical intensity fluctuations. Because $f_{D}$ is near a resonance in H-KPFM, we assume Brownian motion is the dominant force noise. In this limit, the total noise in the signal is:
	\begin{align}
	\label{eq:noise}
	N(B) &= \frac{1}{2}\int^{B}_{-B}\bigg[\frac{2\gamma_{i}^{2} k_{i}k_{B}T}{\pi f_{i}Q_{i}}|G(f_{i}+f)|^{2}\\&+n^{2}_{\text{det}}(f)+n^{2}_{\text{dis}}(f)\bigg]df, \notag
	\end{align}
	where the first term in the brackets represents the noise due to Brownian motion of the cantilever\cite{Heer1972}, $k_{B}$ is Boltzmann's constant, $T$ is temperature, $n_{\text{det}}$ is the detection noise amplitude spectral density (which is nearly constant over the integral), and $n_{\text{dis}}$ is displacement noise amplitude spectral density (which depends on the specifics of KPFM operation). If we consider only the Brownian motion of the cantilever, and assume the detection bandwidth $B$ is less than the bandwidth of the cantilever ($B < f_{i}/2Q_{i}$), then the integral in equation \ref{eq:noise} can be computed analytically:
	\begin{align}
	N(B) &\approx \frac{\gamma_{i}^{2} k_{B}T}{\pi k_{i}}\text{arctan}\bigg(\frac{2Q_{i}B}{f_{i}}\bigg),
	\label{eq:just_Brownian}	
	\end{align}
	yielding the same noise as used in previous calculations of $V_{m}$, in the limit of small $B$\cite{Nonnenmacher1991}.
	
To understand how cantilever characteristics affect the minimum detectable voltage, each eigenmode of $G(f)$ is modeled as a point mass harmonic oscillator\cite{Melcher2007}. Then the minimum detectable voltage becomes: 
\begin{align}
V_{m,\textrm{H}}&= \frac{2\sqrt{2k_{B}T}}{\sqrt{\pi}A_{CA} V_{AC} C''}\sqrt{k_{i}\text{arctan}\bigg(\frac{2Q_{i}B}{f_{i}}\bigg)}.
\label{eq:vm_h1_thermal_smallBlimit} 	
\end{align}
Conversely, if the dominant noise source is broadband detector noise ({\it e.g.} off resonance), then $N(B)\approx n_{\text{det}}^{2} B$. The minimum detectable voltage when detector noise dominates is:
\begin{align}
V_{m,\textrm{H}} = \frac{2 n_{\text{det}}\sqrt{B}}{A_{CA} V_{AC} C''}\frac{1}{|G(f_{D})|\gamma(f_{D})}.
\label{eq:vm_H_detector}
\end{align}
Note that the optical lever sensitivity depends on the eigenmode excited (a cantilever bends more for the same $z$ displacement if excited at higher eigenmodes\cite{Butt1995}).

		\begin{figure}[h]
			\centering
			\includegraphics[width=.45\textwidth]{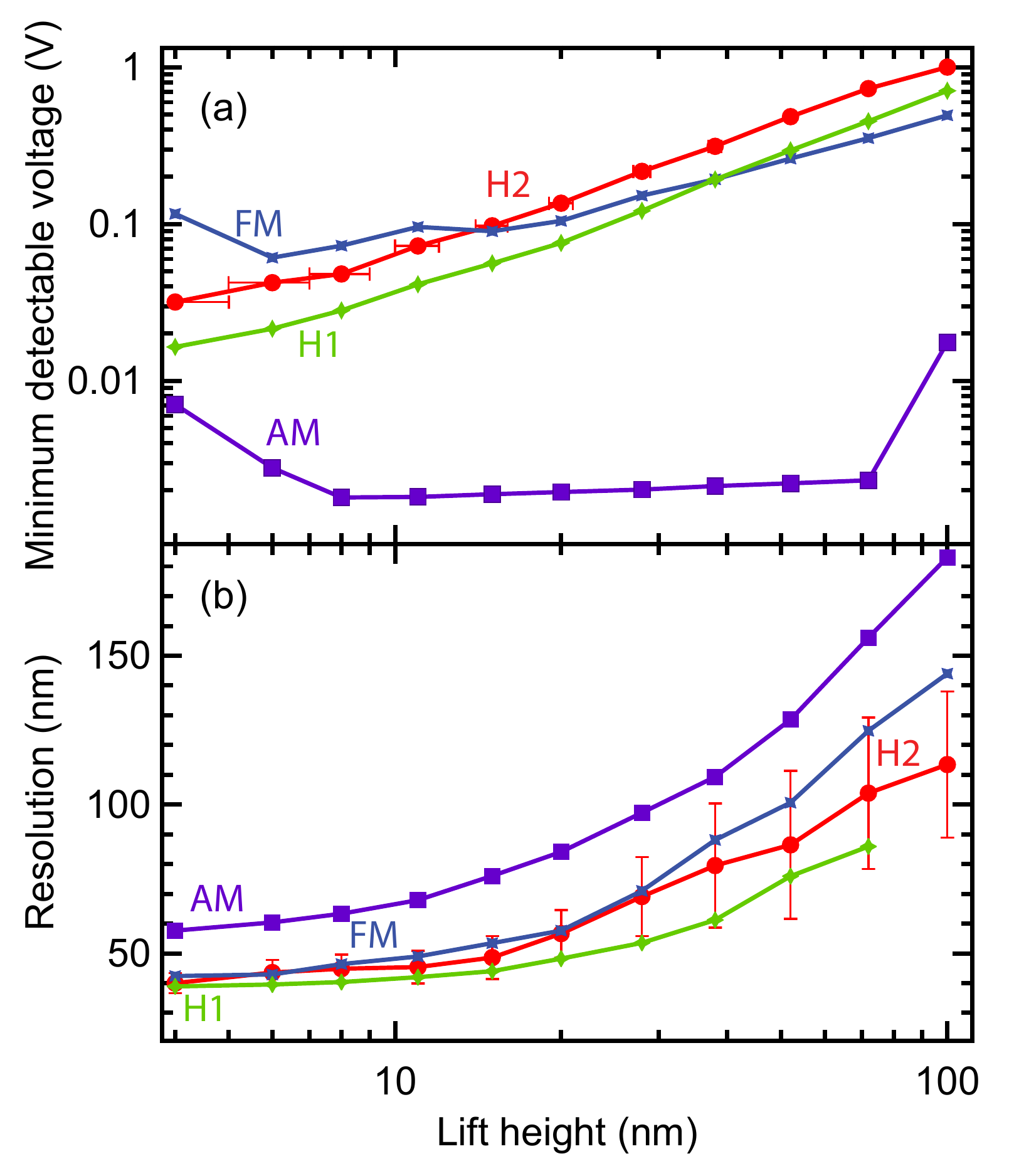}
			\caption{The (a) minimum detectable voltage and (b) 10-90 resolution both increase with lift height, for all methods. The data plotted here are for cantilever topographical oscillation of $A_{CA}=8$ nm. Both heterodyne methods achieve resolutions similar to FM-KFPM.}
			\label{fig:Lift_Height_2}
		\end{figure}
The minimum detectable voltage, $V_{m}$ is experimentally determined by measuring the signals at the lock-in amplifier,  $S_{LIA}^{i}$ and $S_{LIA}^{q}$ (equation \ref{eq:lock_in_signals}), with the feedback loop open. The detection phase, $\phi_{D}$, is swept from -180$^\circ$ to 180$^\circ$ at $V_{K}$ = -1, -0.3, 0.3, and 1 V. For each $\phi_{D}$, the sensitivity $\zeta_{j}$ is determined by fitting $S_{LIA}^{i}$ vs $(V_{K}+V_{0})$ to a line, the slope of which is $\zeta_{j}$ (equation \ref{eq:KPFM_feedback}). Calculating $\zeta_{j}$ for several  $\phi_{D}$, allows us to account for a small systematic offset on the output of the LIA, and to determine the $\phi_{D}$ that maximizes $\zeta_{j}$. The noise at the output of the LIA is sampled at 5 kHz, and the calculations here consider the noise within a bandwidth of 200 Hz. Then equation \ref{eq:Vm} is used to calculate $V_{m}$.

		\begin{figure}[h]
			\centering
			\includegraphics[width=.45\textwidth]{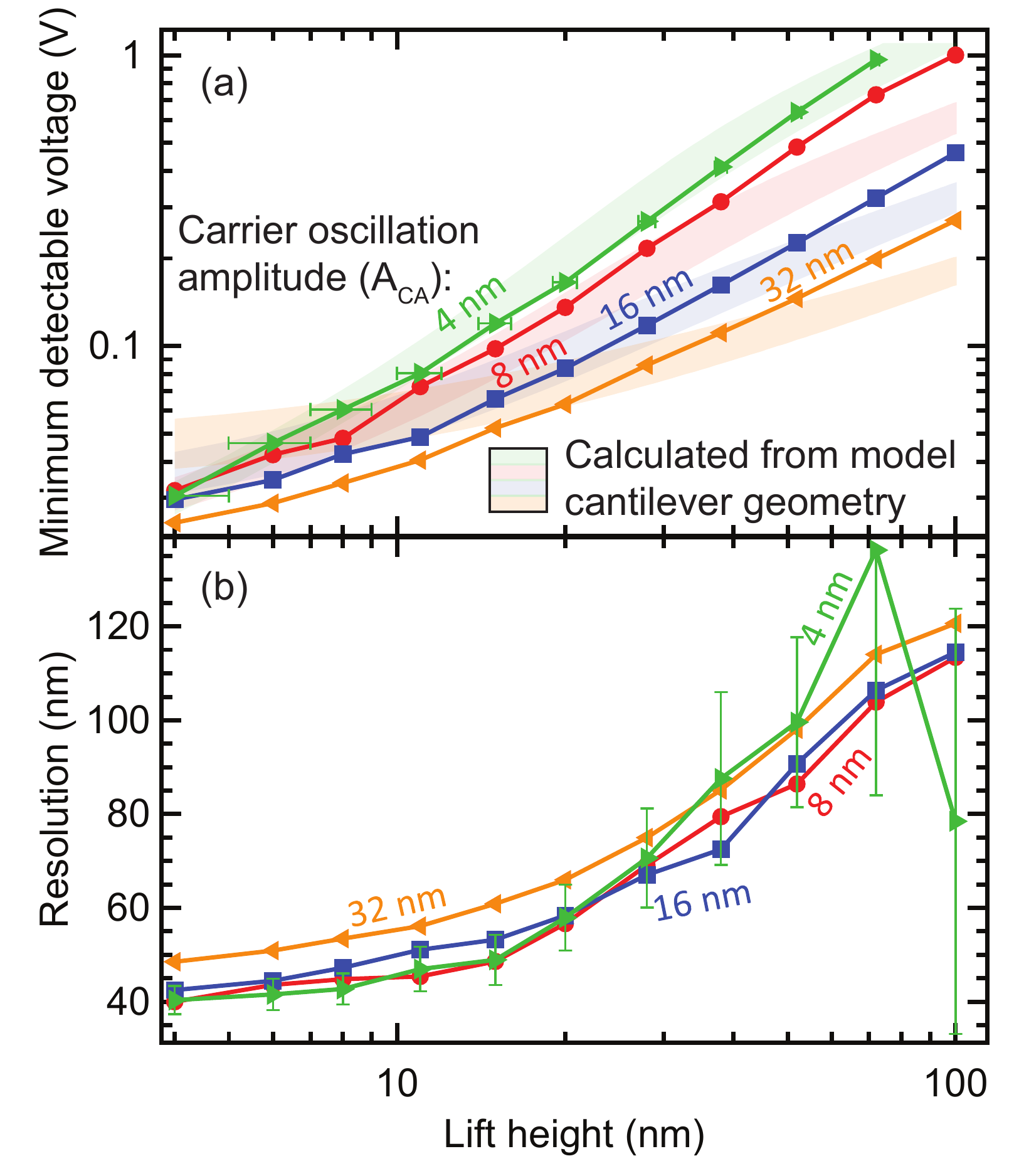}
			\caption{With H-KPFM in the H2 implementation, (a) the minimum detectable voltage, $V_{m}$, and (b) the 10-90 resolution, $l_{10-90}$, both increase with lift height. Larger shake amplitude, $A_{CA}$, decreases $V_{m}$, but no effect on resolution is found { above the noise level} as a function of $A_{CA}$. {A cantilever model is used to calculate expected $V_{m}$ (shaded regions).}
			}
			\label{fig:Lift_Height_1}
		\end{figure} 
	The lift-height dependence of $V_{m}$ for FM- and H-KPFM with different heights is measured. For each lift height, a force curve is used to set the position at the chosen lift height, where the probe is held for the duration of the $V_{m}$ measurement. As the separation is increased, $V_{m}$ increases, for all implementations (figure \ref{fig:Lift_Height_2}a). AM-KPFM has the smallest minimum detectable voltage; however, the small $V_{m}$ is a consequence of the stray capacitance of the cantilever, which causes potential contrast to only be qualitative, and limits spatial resolution\cite{Jacobs1997,Zerweck2005}. Within H2, $V_{m}$ increases more quickly with lift height for smaller $A_{CA}$ (figure \ref{fig:Lift_Height_1}a). {In addition, $V_{m}$ is calculated from a model cantilever geometry\cite{Colchero2001} combined with noise from equation \ref{eq:just_Brownian} for the cantilever described in table \ref{table:cantilevers}, where the tip radius and opening angle are the only free parameters. A tip radius of $16\pm2$ nm with an opening angle of $40\pm5^{\circ}$ is found to approximate the $A_{CA}=4$ nm data. The calculated $V_{m}$ for this geometry, for all $A_{CA}$ are plotted in figure \ref{fig:Lift_Height_1}.}
	
	\begin{figure}[h]
		\centering
		\includegraphics[width=.45\textwidth]{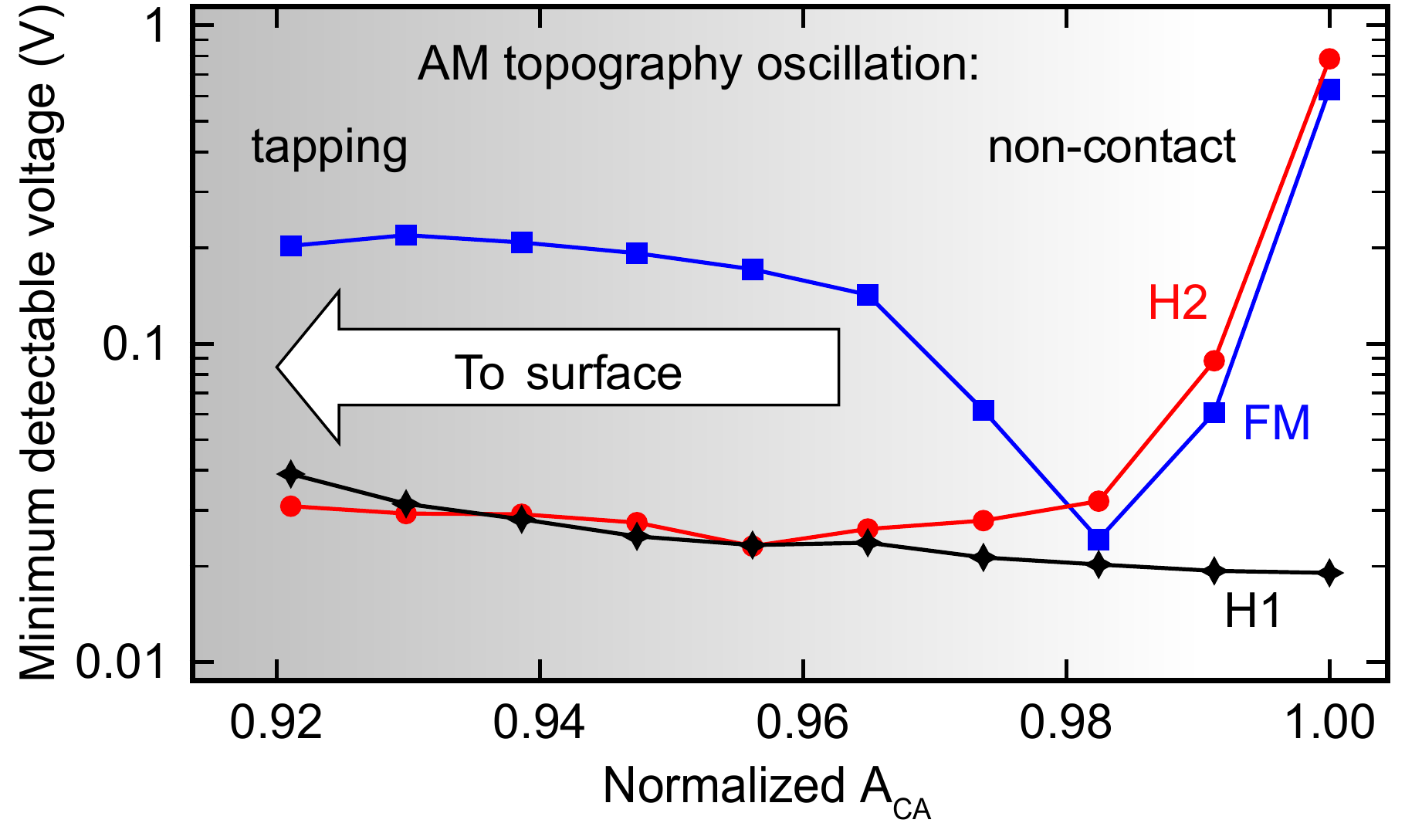}
		\caption{ The minimum detectable voltage changes as a function of the  normalized {$A_{CA}$}, which decreases as the probe moves closer to the surface (as in figure \ref{fig:Distance_curves}). A smaller setpoint moves the cantilever closer to the surface. Far from the surface H2 and FM-KFPM have similar $V_{m}$, but it becomes much greater for FM-KPFM nearer the surface, where the noise increases. The H1 method uses a different eigenmode for topography, and does not have the steep increase in $V_{m}$. The gradient from dark to light represents the change from tapping mode to a non-contact mode as the topography setpoint is increased and the probe is lifted from the surface. }
		\label{fig:Vm_contact}
	\end{figure}
	
	Similarly, we measure $V_{m}$ while in tapping mode, as the topographical setpoint is gradually decreased. The noise in both H-KPFM implementations increases slowly as the setpoint is decreased, but the noise density in FM-KPFM increases rapidly, so that close to the surface, $V_{m}$ for FM-KPFM is about an order of magnitude larger (figure \ref{fig:Vm_contact}). {Because the noise does not increase as rapidly for H1, the source of the noise is not solely due to using the first resonance for KPFM detection. Likewise, because the rapid noise increase is not seen in H2, the source of the extra noise is not solely due to which resonance is used for topography control. Thus, we suspect that the rapid increase in noise when FM-KPFM approaches the surface is due to signal detection} ($f_{D}$) and {topography control ($f_{CA}$)} utilizing the same eigenmode.

\subsection{Spatial resolution}

	 Determining the spatial resolution of KPFM typically involves observing potential change around a boundary. Jacobs {\it et al.} showed that the boundary between two micron-scale objects allows for a clear empirical definition of spatial resolution and calculated a 25-75 resolution, {\it i.e.} the distance over which 50\% of the total observed voltage change occurred, as a function of lift height\cite{Jacobs1997,Jacobs1998,McMurray2002}. Zerweck {\it et al.} similarly calculate a 10-90 resolution\cite{Zerweck2005}. An equation for the resolution from a point probe is derived in \cite{McMurray2002}. Others have sought information about the resolution by comparing the boundaries to particular functions, such as arctangent\cite{McMurray2002} or Boltzmann functions\cite{Liscio2006}.

\begin{figure}[h]
			\centering
			\includegraphics[width=.45\textwidth]{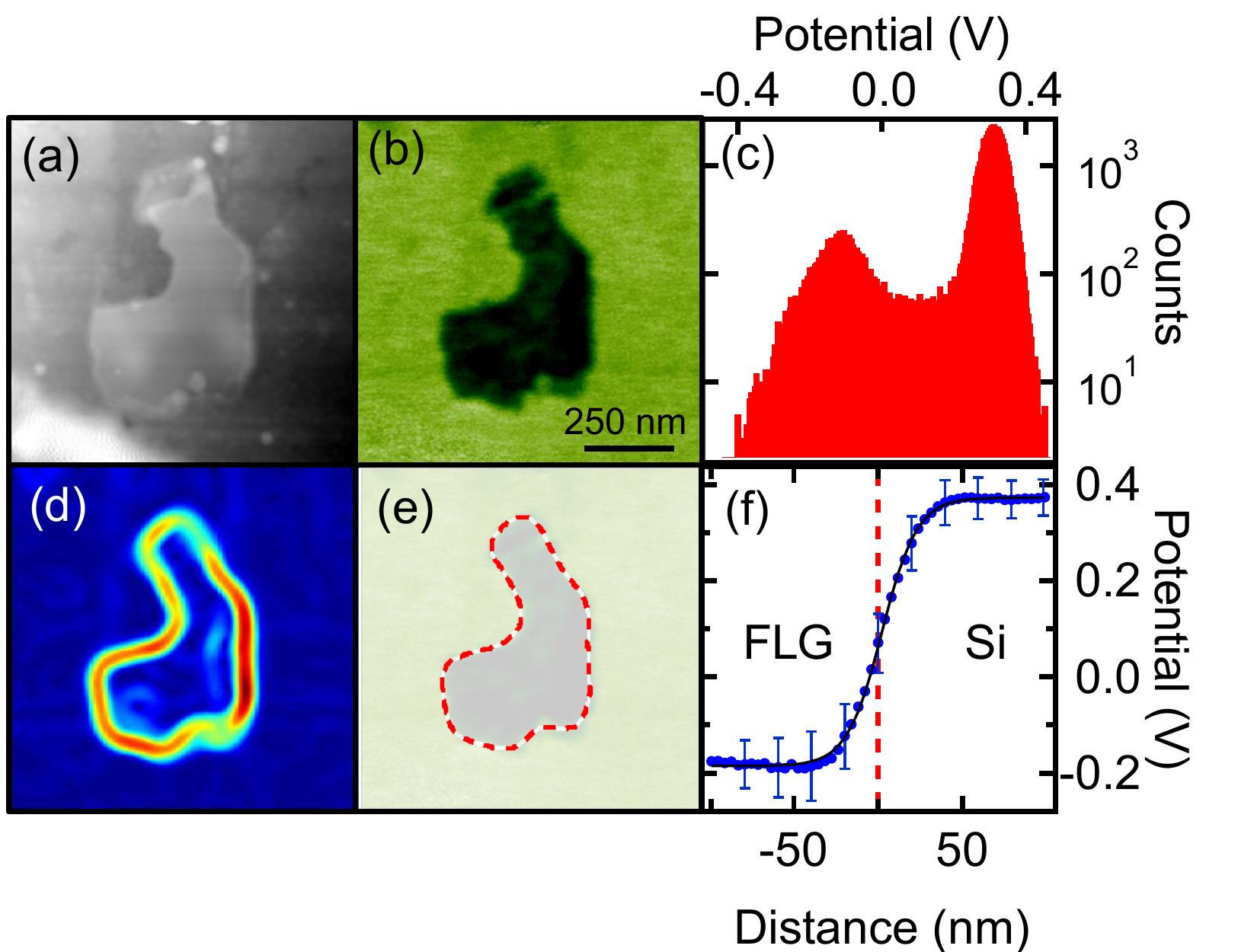}
			\caption{ Few layer graphene on silicon shows (a) height contrast and (b)  significant voltage contrast. (c) A histogram of the KPFM dat shows that the potential distribution is bimodal. (d,e) A watershed algorithm is applied to gradient magnitude of the potential image (d) in order to calculate the boundary (e, {\color{red}$\mathbf{- - -}$}). (f) Voltages are summed as a function of the distance from the boundary ({\color{blue}$\bullet$}) and fit to a tanh function (black solid line), from which the 10-90 resolution is deduced. 
			} 
			\label{fig:Determining_resolution}
		\end{figure}

	Here we estimate a 10-90 resolution, $l_{10\text{-}90}$, by fitting the measured potential as a function of distance from the boundary to a hyperbolic tangent (figure \ref{fig:Determining_resolution}). The theoretically expected form of the measured potential near the boundary is very nearly a tanh within the proximity force approximation, as shown in Appendix \ref{sec:spatial_res_eq}. For large lift heights ($> 20$ nm), the resolution is large enough to prevent $V_{K}$ from reaching its asymptotic value over the scan size, which necessitates the use of a fit. The noise inherent in KPFM is overcome by averaging around the boundary. The equation of the hyperbolic tangent fit to the boundary is:
	\begin{align}
		\overline{V_{K}}(x)=(V_{b} \tanh\big[\text{ln}(9)(x-x_{0})/l_{10\text{-}90}\big] + c)/2,
		\label{eq:arctanfit}
	\end{align}	
	where $V_{b}$ is the potential change across the boundary, $ \overline{V_{K}}(x)$ is the average measured $V_{K}$ a distance $x$ from the boundary, $x_{0}$ is the center of the boundary, and $l_{10\text{-}90}$ is the 10-90 resolution. This fit gives the empirical spatial resolution. In order to determine whether or not the measured potential on either side of the boundary corresponds to the actual potential difference, one must supplement this data with either theory\cite{Zerweck2005} or knowledge of the accuracy of the detected voltage (as in figure \ref{fig:Distance_and_dv}). 
		
	Regions of few layer graphene and silicon are identified by watershed segmentation\cite{Gonzalez2004}. First, the image is median filtered in order to mitigate the effect of noise on the algorithm, and the trace and retrace are averaged. Second, the gradient magnitude of the resultant potential image is calculated with a Sobel algorithm\cite{Gonzalez2004}. Third, points of lowest and highest potential across the image are marked. Fourth, the watershed algorithm is applied with the two marked points forming the origin of each basin (figure \ref{fig:Determining_resolution}). 
	
	Once the image is divided into two components, we plot the potential of the unaltered measurement as a function of the distance from the estimated boundary, and fit the resulting curve to an tanh function (figure \ref{fig:Determining_resolution}e,f). The 10-90 resolution, $l_{10\text{-}90}$, is then extracted from the fit.
		
	For all KPFM methods used, $l_{10\text{-}90}$ increases with lift height (figures \ref{fig:Lift_Height_2},\ref{fig:Lift_Height_1}b), as observed before with AM-KPFM\cite{Jacobs1997}. Both implementations of H-KPFM and FM-KPFM achieve better spatial resolution than AM-KPFM, at all heights, but the error is too large to discern a difference between the former three. However, when the resolution approaches the length of the few layer graphene, or when the minimum detectable voltage reaches the contrast between the objects, the error grows large. Finding a longer, straighter boundary to measure, with larger contrast, could aid in future measurements of resolution at larger lift heights.  
	
\section{Conclusion}

	In this paper, we explore the versatility of H-KPFM and uncover its beneficial characteristics, the most prominent of which is its speed. The H1 implementation improves the minimum detectable voltage by $\approx80\%$ relative to the original implementation.  Further studies into the technique of H-KPFM should investigate the effect of roughness, the effect of eigenmode shape (reportedly an issue with the simpler AM-KPFM\cite{Satzinger2012}), and how to incorporate better control techniques for potential estimation ({\it e.g.} \cite{Wagner2015a}) and tracking of the surface({\it e.g.} \cite{Ahmad2014}), which now limits KPFM scan speed. Cantilevers could be designed specifically for H-KPFM\cite{Sadewasser2006} to reduce the difference between the spring constants of the first and second eigenmodes, which would improve the sensitivity of H-KPFM. Likewise, cantilever resonance frequencies could be chosen to enable open-loop H-KPFM\cite{Collins2013}.

	Heterodyne KPFM improves upon the time resolution of FM-KPFM. Rates of several frames per minute are achieved. Its speed is not limited by AC coupling or bandwidth overlap, and so with appropriate cantilevers it will operate even faster. It also improves upon the spatial resolution of AM-KPFM. These new implementations of H-KPFM will facilitate fast and accurate measurements of nanoscale potential dynamics.

	\section{Acknowledgements}
	We thank Asylum Research for technical advice, in particular Anil Gannepalli. We thank Tao Gong, Beth Tennyson, and Marina Leite for insightful conversations about KPFM and for challenging us with difficult-to-scan samples, and Dakang Ma and David Somers for critical readings of this article. We thank the University of Maryland for financial support.

\appendix
\onecolumngrid
\begin{table}[h]
	\center
	\section{Table of Variables}\label{sec:table_of_variables}
	\small
	\begin{tabular}{ c | l | l }
		Variable & Description  &  Equations \\
		\hline
		$S_{K}$ & KPFM signal & \ref{eq:KPFM_feedback},\ref{eq:SKforAM},\ref{eq:Extra_signal} \\ 
		$V_{K}$,$V^{C^{(n+1)}}_{K}$ & KPFM voltage, KPFM voltage near a boundary & \ref{eq:KPFM_feedback},\ref{eq:AM_force},\ref{eq:at_photodetector},\ref{eq:lock_in_signals},\ref{eq:SKforAM},\ref{eq:All_u_terms},\ref{eq:H_KPFM},\ref{eq:signal_H},\ref{eq:extra_voltage_inVK},\ref{eq:calibratin},\ref{eq:V_Kexact},\ref{eq:fits},\ref{eq:adef}	 \\ 
		$V_{0}$ & Inherent contact potential difference & \ref{eq:KPFM_feedback},\ref{eq:AM_force},\ref{eq:at_photodetector},\ref{eq:lock_in_signals},\ref{eq:SKforAM},\ref{eq:All_u_terms},\ref{eq:H_KPFM},\ref{eq:signal_H},\ref{eq:extra_voltage_inVK} \\ 
		$\zeta_{j}$ & Sensitivity of KPFM method $j$, where $j$ = AM, FM, or H & \ref{eq:KPFM_feedback},\ref{eq:SKforAM},\ref{eq:chi_am},\ref{eq:heterodyne_chi},\ref{eq:V_error},\ref{eq:calibratin},\ref{eq:FM_error},\ref{eq:Vm} \\ 
		$f_{D}$ & Frequency at which the KPFM signal is detected & \ref{eq:AM_force},\ref{eq:lock_in_signals},\ref{eq:heterodyne_chi} \\
		$F_{f_{D}} $ & Force on cantilever at detection frequency & \ref{eq:AM_force},\ref{eq:amplitude},\ref{eq:H_KPFM} \\
		$C$ & Tip-sample capacitance & \ref{eq:AM_force},\ref{eq:at_photodetector},\ref{eq:lock_in_signals},\ref{eq:chi_am},\ref{eq:All_u_terms},\ref{eq:H_KPFM},\ref{eq:signal_H},\ref{eq:vm_h1_thermal_smallBlimit},\ref{eq:vm_H_detector} \\
		$V_{AC}$ & Periodic voltage applied to cantilever & \ref{eq:AM_force},\ref{eq:at_photodetector},\ref{eq:chi_am},\ref{eq:All_u_terms},\ref{eq:H_KPFM},\ref{eq:signal_H},\ref{eq:heterodyne_chi},\ref{eq:vm_h1_thermal_smallBlimit},\ref{eq:vm_H_detector}\\
		$A_{f_{D}}$& Amplitude of cantilever oscillation at $f_{D}$&\ref{eq:amplitude}\\
		$G(f)$ & Transfer function of cantilever at frequency $f$&\ref{eq:amplitude},\ref{eq:at_photodetector},\ref{eq:lock_in_signals},\ref{eq:chi_am},\ref{eq:signal_H},\ref{eq:heterodyne_chi},\ref{eq:vm_H_detector}\\
		$S_{\text{photo}}$ & KPFM signal at photodetector &\ref{eq:at_photodetector},\ref{eq:signal_H}\\
		$\gamma(f),\gamma_{i}$ & Optical lever sensitivity of cantilever at frequency $f$ or eigenmode $i$ &\ref{eq:at_photodetector},\ref{eq:lock_in_signals},\ref{eq:chi_am},\ref{eq:signal_H},\ref{eq:heterodyne_chi},\ref{eq:FM_error},\ref{eq:noise},\ref{eq:just_Brownian} \\
		$S_{LIA}^{i}$,$S_{LIA}^{q}$& In-phase ($i$) and $\pi/2$-shifted ($q$) signals at lock-in &\ref{eq:lock_in_signals},\ref{eq:SKforAM},\ref{eq:Extra_signal}\\
		$\phi_{D}$ & Phase of shift of lock-in amplifier & \ref{eq:lock_in_signals},\ref{eq:chi_am},\ref{eq:heterodyne_chi} \\
		$z$, ($\bar{z}$) & Instantaneous (time-averaged) tip-sample separation &\ref{eq:expand_z},\ref{eq:calibratin},\ref{eq:PFA},\ref{eq:force_der},\ref{eq:lambda_def},\ref{eq:cpres},\ref{eq:resolutions}\\
		$F$ & Capacitive force on cantilever & \ref{eq:All_u_terms},\ref{eq:PFA},\ref{eq:force_der}\\
		$t$ & Time &\ref{eq:expand_z},\ref{eq:FM_error}\\
		$A_{CA}$& Amplitude of carrier oscillation, also used for topography feedback&\ref{eq:expand_z},\ref{eq:All_u_terms},\ref{eq:H_KPFM},\ref{eq:signal_H},\ref{eq:heterodyne_chi},\ref{eq:FM_error},\ref{eq:vm_h1_thermal_smallBlimit},\ref{eq:vm_H_detector}\\
		$f_{CA}$ & Frequency of carrier oscillation & \ref{eq:expand_z},\ref{eq:All_u_terms}\\
		$\phi_{CA}$&Phase of carrier oscillation&\ref{eq:expand_z},\ref{eq:heterodyne_chi}\\
		$f_{A}$ & Frequency at which $V_{AC}$ is applied to the cantilever &\ref{eq:All_u_terms},\ref{eq:FM_error} \\
		$\phi_{A}$&Phase of applied voltage $V_{AC}$&\ref{eq:All_u_terms},\ref{eq:heterodyne_chi}\\
		$S_{E}$ & Extraneous signal in KPFM feedback & \ref{eq:Extra_signal},\ref{eq:V_error},\ref{eq:calibratin}\\
		$V_{E}{,V_{E}^{CA}}$ & Extraneous voltage artifact & \ref{eq:extra_voltage_inVK},\ref{eq:V_error},\ref{eq:calibratin},\ref{eq:FM_error}\\
		$S_{sp}$ & KPFM feedback setpoint & \ref{eq:V_error},\ref{eq:calibratin}\\
		$B$ & Bandwidth of low-pass filter on lock-in amplifier & \ref{eq:FM_error},\ref{eq:Vm},\ref{eq:noise},\ref{eq:just_Brownian},\ref{eq:vm_h1_thermal_smallBlimit},\ref{eq:vm_H_detector}\\
		$V_{p}$ & Voltage perturbation applied to plate & \\
		$f_{p}$ & Frequency of voltage perturbations applied to plate&\\
		$t_{c}=1/f_{c}$ & The cut-off time and the cut-off frequency&\\
		$V_{m}$ & Minimum detectable voltage & \ref{eq:Vm},\ref{eq:vm_h1_thermal_smallBlimit},\ref{eq:vm_H_detector}\\
		$N(B)$ & Noise power in detection bandwidth & \ref{eq:Vm},\ref{eq:noise},\ref{eq:just_Brownian}\\
		$k_{i}$, $Q_{i}$, $f_{i}$ & Spring constant, quality factor, or frequency of eigenmode $i$ & \ref{eq:noise},\ref{eq:just_Brownian},\ref{eq:vm_h1_thermal_smallBlimit}\\
		$n_{\text{det}},n_{\text{dis}}$& Detection and displacement noise amplitude spectral densities &\ref{eq:noise},\ref{eq:vm_H_detector} \\
		$V_{b}$ & Surface voltage change across a boundary & \ref{eq:arctanfit},\ref{eq:V_Kexact},\ref{eq:am_res},\ref{eq:fits} \\
		$L\equiv x-x_{0}$ & Distance from potential boundary & \ref{eq:arctanfit},\ref{eq:lambda_def},\ref{eq:V_Kexact},\ref{eq:fits},\ref{eq:adef} \\
		$l_{10-90}$,$l_{10-90}^{C^{(n+1)}}$  & 10-90 resolution, for method that depends on $C^{(n+1)}$ & \ref{eq:arctanfit},\ref{eq:cpres},\ref{eq:resolutions}\\
		$R$ & Tip radius & \ref{eq:PFA},\ref{eq:force_der},\ref{eq:lambda_def},\ref{eq:cpres},\ref{eq:resolutions}\\
		$V_{pl}(r,\phi)$, $V$ & Surface potential of the plate, sphere in PFA & \ref{eq:PFA},\ref{eq:force_der},\ref{eq:lambda_def}\\
		\hline
	\end{tabular}
	\label{table:variables}
\end{table}	

\FloatBarrier

\twocolumngrid
	\section{An equation for spatial resolution}\label{sec:spatial_res_eq}
	Above we discuss the spatial resolution of KPFM in terms of $l_{10\text{-}90}$, the 10-90 resolution, or the distance over which 80\% of the voltage change across a boundary occurs. We determine $l_{10\text{-}90}$ by fitting $V_{K}(x)$, the potential measured across a boundary, to a hyperbolic tangent (equation \ref{eq:arctanfit}).
	
	Here, we use the proximity force approximation (PFA) for a sphere interacting with a plate to derive an analytic expression for $V_{K}(x)$ for both $C'$ and $C''$ KPFM methods. Furthermore, we demonstrate that the tanh function approximates the form of $V_{K}(x)$ better than the arctan function in order to motivate our choices in the text. Finally, we estimate how $l_{10\text{-}90}$ changes with height and tip radius. We note that an equation for resolution exists in the large separation, small probe limit\cite{McMurray2002}, but better resolution is achieved with small tip-sample separation, and so that is our focus here.
	
	The PFA for the capacitive force of a sphere above a plate can be written as\cite{Kim2010}:
	\begin{align}
	F(z) = \frac{\epsilon_{0}}{2} \int_{0}^{2\pi}{d\phi}\int_{0}^{R}r dr \frac{(V-V_{pl}(r,\phi))^2}{(z + r^{2}/2R)^2},
	\label{eq:PFA}
	\end{align}
	where $R$ is the radius of the sphere, $V_{p}(r,\phi)$ is the potential of the plate at position $(r,\phi)$, and $V$ is the potential of the sphere (here assumed to be spatially uniform). 
	The voltage applied to the probe that minimizes $n$th derivative of this force can be found by taking $n$ derivatives with respect to $z$ and one with respect to $V$,
	\begin{align}
	\frac{\partial^{n+1}F(z)}{\partial V\partial z^{n}} = \frac{\partial}{\partial z^{n}}\Biggl[\epsilon_{0} \int_{0}^{2\pi}{d\phi}\int_{0}^{R}r dr \frac{V-V_{pl}(r,\phi)}{(z + r^{2}/2R)^2}\Biggr].
	\label{eq:force_der}
	\end{align}
	At the Kelvin probe voltage, $V=V_{K}(x)$, for which the KPFM signal $S_{K}$ vanishes, equation \ref{eq:force_der} vanishes as well. Near a boundary, the potential of the plate is $V_{pl}(r,\phi) = V_{b}\Theta(\cos(\phi)+L/r)$, where $L=x-x_{0}$ is the distance between the location of the probe and the boundary and $\Theta$ is the Heaviside step function . The potential is $V_{b}$ for $r\cos(\phi)>-L$ and 0 otherwise. To simplify the calculation, we define the function $\Lambda(z,R,L)$:
	\begin{gather}
	\label{eq:lambda_def}
	\Lambda(z,R,L)=\int_{0}^{2\pi}{d\phi}\int_{0}^{R}r dr \frac{\Theta(\cos(\phi)+L/r)}{(z + r^{2}/2R)^2},\\
		= R \left(\frac{2 L \text{ arctan}\left(\sqrt{\frac{R^2-L^2}{L^2+2 R z}}\right)}{z \sqrt{L^2+2 R z}}+\frac{2 R \text{ arccos}\left(-\frac{L}{R}\right)}{z (R+2z)}\right).\notag
	\end{gather}
	For a KPFM method with a signal proportional to the $(n+1)^{\text{th}}$ derivative of capacitance, the Kelvin probe voltage near a boundary is: 
	\begin{align}
	V^{C^{(n+1)}}_{K}(L)=V_{b}\frac{\Lambda^{(n)}}{2\Lambda^{(n)}|_{L=0}},
	\label{eq:V_Kexact}
	\end{align}
	where $\Lambda^{(n)}=\frac{\partial^{n}\Lambda}{\partial z^{n}}$, and $C^{(n+1)}$ represents a method that depends on the $(n+1)^{\text{th}}$ derivative of capacitance. For example, for AM-KPFM, the signal of which is proportional to $C'$, the Kelvin probe voltage is:
	\begin{align}
	V^{C^{(1)}}_{K}(L)=&V_{b}\bigg[\frac{1}{2}+\frac{\text{ arcsin}\left(\frac{L}{R}\right)}{\pi}\notag \\&+\frac{L (2 z+R) \text{ arctan} \left(\sqrt{\frac{R^2-L^2}{L^2+2 z R}}\right)}{\pi  R \sqrt{2 z R+L^2}}\bigg].
	\label{eq:am_res}
	\end{align}
	It must be noted that the PFA only considers the contribution of the tip apex to the KPFM signal. In AM-KPFM the dominant contribution to the signal comes from the cantilever. At a boundary, equation \ref{eq:am_res} will predict the shape of $V^{C^{(1)}}_{K}(x)$, but the $V_{b}$ coefficient will be much less than the potential difference across the boundary. AM-KPFM measures qualitative potential contrast\cite{Jacobs1997}.	
	
\begin{figure}[h]
		\centering
		\includegraphics[width=0.45\textwidth]{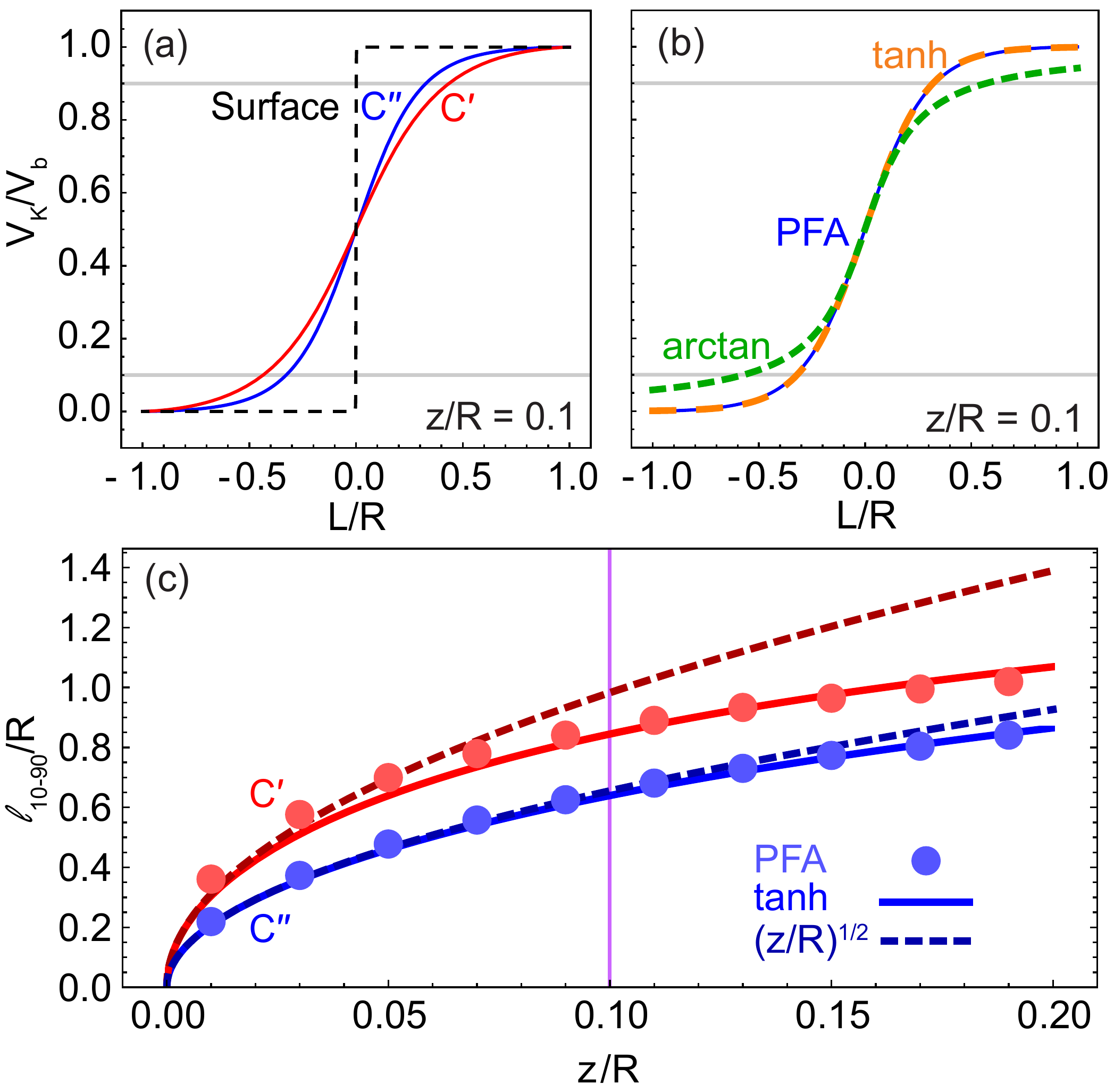}
		\caption{ The normalized Kelvin probe voltage near a boundary in the proximity force approximation (PFA) is shown in (a) for both the AM-KPFM ($C'$, red) and H- and FM-KPFM ($C''$, blue) variants. The dashed black line shows the normalized potential on the surface directly and two grey lines indicate the 10-90 potential change. (b) A tanh function (dashed orange) is a much better fit to the analytic expression (PFA) for the $C''$ method than an arctan function chosen in the same way (dashed green). (c) The 10-90 resolution predicted by a tanh function fit (solid) is compared to that calculated numerically by the exact PFA expression (${\color{blue}\bullet}$). At small z/R, the 10-90 resolution increases with separation $\propto \sqrt{z/R}$, and this approximate expression is plotted for both the $C'$ and $C''$ methods (dashed). A purple line indicates the value of $z/R$ for which (a,b) are plotted.
		}
		\label{fig:Appendix}
	\end{figure}

	For the $C''$ methods (H- or FM-KPFM) the minimizing potential equation is more complicated, and so it has been plotted in figure \ref{fig:Appendix}a.
	 To facilitate data analysis, a simpler function can be used to approximate equation \ref{eq:V_Kexact}. Both arctan and tanh functions have the desired behavior: monotonic, odd around $L=0$, and asymptotic to a constant as $L\rightarrow\infty$. The slope of $V^{C^{(n+1)}}_{K}(L)$ is steepest at $L=0$ and so fitting for small $L$ is most important. Arctan and tanh are used to approximate equation \ref{eq:V_Kexact} by matching the first derivative of each function to our exact analytic expression:
	 \begin{align}
	 \label{eq:fits}
	 V^{C^{(n+1)}}_{\text{tanh}}=& V_{b}\frac{\tanh(2 a^{C^{(n+1)}} L)+1}{2},\\
	 V^{C^{(n+1)}}_{\text{arctan}}=& V_{b}\frac{\arctan(\pi a^{C^{(n+1)}} L)+\pi/2}{\pi}, \notag	 
	 \end{align}
	where, 
	\begin{align}
	a^{C^{(n+1)}}=& \frac{\partial V^{C^{(n+1)}}_{K}}{\partial L}\Bigg|_{L=0},	 
	\label{eq:adef}
	\end{align}
	Both functions are plotted in figure \ref{fig:Appendix}b to visually depict how well each fits equation \ref{eq:V_Kexact}. The tanh fit follows the exact expression more closely than the arctan fit. The tanh fit can then be used to estimate $l_{10-90}$ as a function of z and R:
	\begin{align}
	 l_{10-90}^{C^{(n+1)}} =& \frac{\text{ln}(9)}{2a^{C^{(n+1)}}}.  
	\end{align}
	Which, for AM-KFPM is:
	\begin{align}
	 	l_{10-90}^{C^{(1)}}=&\frac{\pi  \log (81) \sqrt{zR}}{2 \sqrt{z /R}+\sqrt{2} (2 z/R+1) \text{ arctan}\left(\sqrt{\frac{R}{2z}}\right)}.
	 	\label{eq:cpres}
	\end{align}
	
	The more complicated expression of $l_{10-90}^{C''}$ is plotted in figure \ref{fig:Appendix}c. Taylor expanding around $\sqrt{z/R}\approx0$, the resolutions are: 
		\begin{align}
		\label{eq:resolutions}
		\frac{l_{10-90}^{C^{(1)}}}{R} \approx \sqrt{2}\text{ ln}(9)\sqrt{\frac{z}{R}} &\text{  (AM method)},\\
		\frac{l_{10-90}^{C^{(2)}}}{R} \approx \frac{2}{3}\sqrt{2}\text{ ln}(9)\sqrt{\frac{z}{R}} &\text{  (H-,FM- methods)}. \notag
		\end{align}
		Jump-to-contact limits how small z can become, and consequently limits the possible spatial resolution. These approximations are also compared to the exact PFA result in figure \ref{fig:Appendix}c. 
		
		The resolutions calculated here are a lower bound on the resolution possible with KPFM because many components of the probe that would broaden the resolution are neglected. Though the electrostatic probe-surface force from the tip cone and cantilever have been calculated for uniform potential\cite{Hudlet1998,Colchero2001}, we are unaware of any analytic procedure to take into account variations of the surface potential. A procedure does exist to calculate the electrostatic force between a sphere and a plate with potential variations\cite{Behunin2012a}, but the KPFM probe geometry is only slightly better represented by such a model. Most importantly, these extra cases all reduce to the PFA near the surface, where the best spatial resolution is achieved.

\bibliographystyle{iopart_num_edited}
\bibliography{HeterodynePaper}

\end{document}